\let\TeXyear\year
\NewCommandCopy{\LaTeXtextbf}{\textbf}
\documentclass{ieeeaccess}
\RenewCommandCopy{\textbf}{\LaTeXtextbf}
\usepackage{cite}
\usepackage{amsmath,amssymb,amsfonts}
\usepackage{algorithmic}
\usepackage{graphicx}
\usepackage{textcomp}
\usepackage[table,xcdraw]{xcolor} 
\usepackage{color,soul}
\usepackage[]{hyperref}
\usepackage{footmisc}
\usepackage{hyperref}
\hypersetup{
    colorlinks,
    linkcolor={red!50!black},
    citecolor={blue!50!black},
    urlcolor={blue!80!black}
}
\usepackage{wasysym}
\usepackage{xfrac}
\usepackage{scalerel}
\usepackage[algoruled,lined,linesnumbered,commentsnumbered]{algorithm2e}
\usepackage{amsthm,thmtools}
\usepackage{caption}
\usepackage{subcaption}

\newcommand{\eg}{{\em e.g.,} }
\newcommand{\ie}{{\em i.e.,} }

\newtheorem{definition}{Definition}
\newtheorem{theorem}{Theorem}
\newtheorem{lemma}{Lemma}

\newtheorem{innercustomthm}{Theorem}
\newenvironment{customthm}[1]
  {\renewcommand\theinnercustomthm{#1}\innercustomthm}
  {\endinnercustomthm}

\newtheorem{innercustomlemma}{Lemma}
\newenvironment{customlemma}[1]
  {\renewcommand\theinnercustomlemma{#1}\innercustomlemma}
  {\endinnercustomlemma}

\newcommand{\Conv}{%
  \mathop{\scalebox{1.5}{\raisebox{-0.2ex}{$\circledast$}}
  }
}

\newcommand{\footnoteref}[1]{\textsuperscript{\ref{#1}}}

\let\ieeeaccessyear\year
\let\year\TeXyear
\usepackage{tikz}
\usetikzlibrary{automata, positioning, arrows}
\let\year\ieeeaccessyear
\definecolor{accessblue}{RGB}{0,105,154}
\sethlcolor{white!40}

\usepackage{bm}
\makeatletter
\AtBeginDocument{\DeclareMathVersion{bold}
\SetSymbolFont{operators}{bold}{T1}{times}{b}{n}
\SetSymbolFont{NewLetters}{bold}{T1}{times}{b}{it}
\SetMathAlphabet{\mathrm}{bold}{T1}{times}{b}{n}
\SetMathAlphabet{\mathit}{bold}{T1}{times}{b}{it}
\SetMathAlphabet{\mathbf}{bold}{T1}{times}{b}{n}
\SetMathAlphabet{\mathtt}{bold}{OT1}{pcr}{b}{n}
\SetSymbolFont{symbols}{bold}{OMS}{cmsy}{b}{n}
\renewcommand\boldmath{\@nomath\boldmath\mathversion{bold}}}
\makeatother

\def\BibTeX{{\rm B\kern-.05em{\sc i\kern-.025em b}\kern-.08em
    T\kern-.1667em\lower.7ex\hbox{E}\kern-.125emX}}

\begin{document}
\history{Date of publication 26 February 2025.}
\doi{10.1109/ACCESS.2025.3546033}

\title{Performance Analysis: Discovering  Semi-Markov Models From Event Logs}
\author{\uppercase{Anna Kalenkova}\authorrefmark{1},
\uppercase{Lewis Mitchell}\authorrefmark{1}, and Matthew Roughan\authorrefmark{1},
\IEEEmembership{Fellow, IEEE}}

\address[1]{Adelaide Data Science Centre (ADSC), School of Computer and Mathematical Sciences, The University of Adelaide, North Terrace Campus, Adelaide, 5000, South Australia, Australia}
\tfootnote{This work was supported by the Australian Government
through the Australian Research Council’s Discovery Projects
funding scheme (project DP210103700).}

\markboth
{Kalenkova \headeretal: Performance Analysis: Discovering Semi-Markov
Models From Event Logs}
{Kalenkova \headeretal: Performance Analysis: Discovering Semi-Markov
Models From Event Logs}

\corresp{Corresponding author: Anna Kalenkova (e-mail: anna.kalenkova@adelaide.edu.au).}

\begin{abstract}
 \emph{Process mining} is a well-established discipline of data analysis focused on the discovery of process models from information systems' event logs. Recently, an emerging subarea of process mining, known as \emph{stochastic process discovery}, has started to evolve. Stochastic process discovery considers frequencies of events in the event data and allows for a more comprehensive analysis. In particular, when the durations of activities are presented in the event log, performance characteristics of the discovered stochastic models can be analyzed, e.g., the \emph{overall process execution time} can be estimated. Existing performance analysis techniques usually discover stochastic process models from event data, and then simulate these models to evaluate their execution times. These methods rely on empirical approaches.  This paper proposes \emph{analytical techniques} for performance analysis that allow for the derivation of statistical characteristics of the overall processes' execution times in the presence of arbitrary time distributions of events modeled by semi-Markov processes. 
 
 The proposed methods include \emph{express analysis}, focused on the mean execution time estimation, and \emph{full analysis} techniques that build probability density functions (PDFs) of process execution times in both continuous and discrete forms. These methods are implemented and tested on real-world event data, demonstrating their potential for \emph{what-if} analysis by providing solutions without resorting to simulation. \hl{Specifically, we demonstrated that the discrete approach is more time-efficient for small duration support sizes compared to the simulation technique. Furthermore, we showed that the continuous approach, with PDFs represented as Mixtures of Gaussian Models (GMMs), facilitates the discovery of more compact and interpretable models.}
\end{abstract}


\titlepgskip=-21pt

\maketitle

\section{Introduction}
\label{sec:intro}

\Figure[th!](topskip=0pt, botskip=0pt, midskip=0pt)[width=0.8\textwidth]{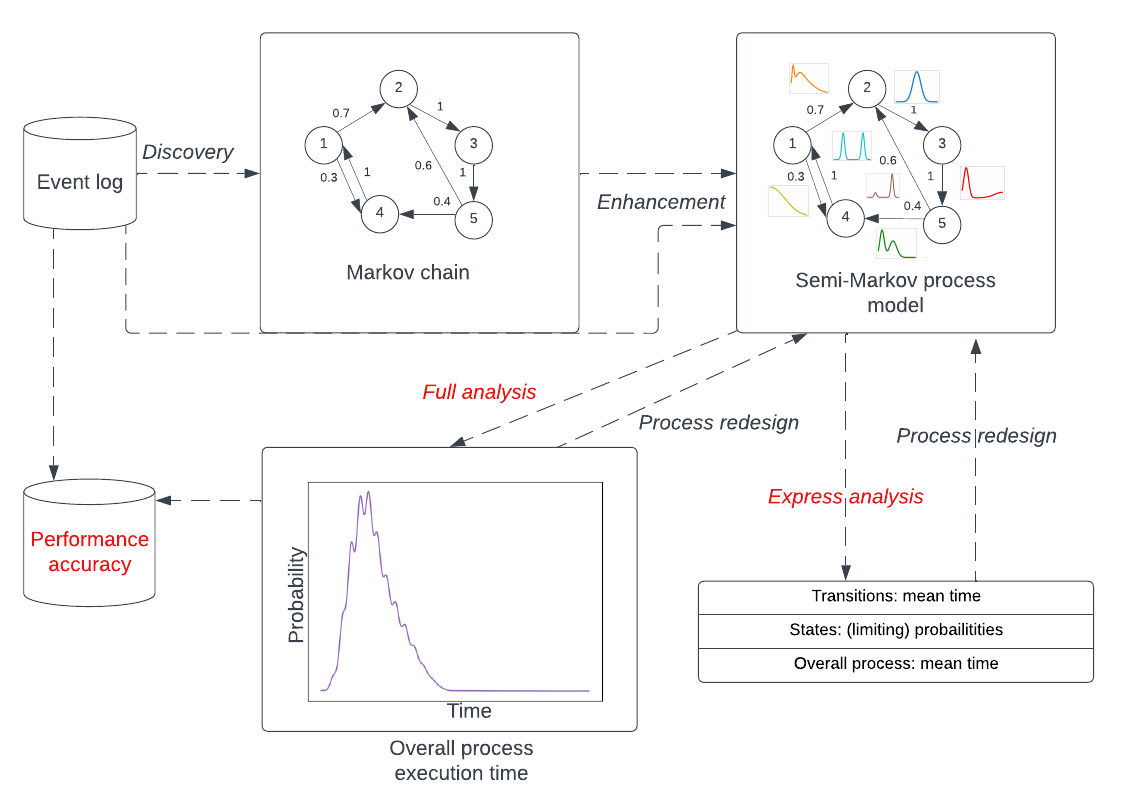}
{ \textbf{Discovery and performance analysis of stochastic process models.}\label{fig:schema}}

\emph{Process mining}~\cite{Aalst16} offers various methods to analyze \hl{event logs from information systems, which record process activities along with their timestamps.} These methods include: (1)~\emph{automatic process discovery}, which aims at learning process models from event logs; these models generalize and visualize process behavior, making it easier for end users to analyze the processes, (2) \emph{conformance checking}, which identifies discrepancies between process models (expected system behavior) and event logs (actual system behavior), and (3) \emph{performance analysis}, which uses event logs to identify bottlenecks as well as assess and predict the overall performance of the system.

This paper integrates automatic process discovery and performance analysis by studying performance characteristics of the discovered process models and relating them to the actual execution times recorded in the event logs.

Among different performance characteristics, \emph{overall process execution time} (the time from start of the process to its completion) is of particular interest, because, if detailed enough, it provides analysts with valuable insight into process execution and can assist in process optimization. Several methods for the analysis of process execution times have been proposed in process mining. However, these methods are either based on simulation of the discovered stochastic models~\cite{solti-predict-simulation,ROGGESOLTI20151,vandenabeele2022enhancing,ROZINAT2009305,CAMARGO2020113284} or predict the remaining time~\cite{VANDERAALST2011450,10.1007/978-3-642-13094-6_5,context-aware} relating the current state of the process to the corresponding state in the event log. The first group of approaches relies on model simulations, which could be non-deterministic and lack interpretability, and the second group of performance analysis techniques cannot be applied if the process model is modified, because these approaches solely rely on event logs. In this paper, we propose \emph{general solutions} for \emph{stochastic performance analysis} without resorting to simulations, and offer tools for further optimizations by allowing modification and analysis of the discovered stochastic process~models. For a detailed consideration of the related work please refer to~\autoref{sec:rel_work}.

Unlike simple \hl{standard continuous} Markov models, \hl{where durations are exponentially distributed}, our approach allows 
%a parsimonious representation of
\emph{arbitrary time distributions}, because, as will be shown in~\autoref{sec:cont_time_dist}, durations of events in real-world data can be \hl{multi-modal and deviate significantly from exponential distributions}.
We represent arbitrary time distributions through semi-Markov models, which provide much of the simplicity of Markov models, but allow much more generality in inter-event times.

Figure~\ref{fig:schema} presents the overall schema of the performance analysis. First, a Markov chain (represented by states and probabilities of transitions between the states) is \emph{discovered}. Then, this Markov chain can be \emph{enhanced} with additional time information, \ie the corresponding semi-Markov model is built. This semi-Markov model additionally considers  a time perspective, represented by arbitrary probability density functions of waiting times associated with the transitions (refer to~\autoref{sec:proces_discovery}).

Once the model is discovered, it can be analyzed to retrieve detailed information on the \emph{overall process execution time}. \emph{Express analysis} (see~\autoref{sec:express_analysis} for the express analysis description and~\autoref{sec:express_case_study} for its application to real-world data) allows one to retrieve the \emph{mean} of the process execution time based on the mean waiting times for each of the transitions. Additionally,  probabilities of visiting the states in a long process run (so-called \emph{limiting probabilities}) are calculated. These probabilities provide essential information on the impact of each of the states on the overall process execution time and can be further used in process redesign.

\emph{Process redesign} answers ``what-if'' questions and can be done in two ways: (1) \emph{control-flow optimization}, which involves changing the structure of the process, \ie transitions between the states and their probabilities are modified; (2) \emph{time optimization}: the waiting times of some states are modified. In other words, the process is either structurally reorganized or the time for performing an activity is changed, for example, the time might be reduced by assigning more resources to that task.

Our \emph {full analysis} (see~\autoref{sec:full_analysis} for its description and ~\autoref{sec:time_dist} for its application to real data) considers both continuous (represented as mixtures of Gaussians, also known as GMMs) and discrete time distributions, and constructs the full probability distribution of the overall process execution time based on a semi-Markov model, which can be obtained as a result of applying process discovery and enhancement algorithms or at a process redesign step. The semi-Markov model approximates and generalizes the event log, hence, the overall execution time of the model can deviate from the \hl{ground truth} overall execution time observed in the log. To quantify this deviation, \emph{performance accuracy} of the model can be estimated (\autoref{sec:case_studies}). This will allow assessment of the quality of the process discovery and enhancement~algorithms. 

The contributions of this paper are as follows: (1) a method for discovering semi-Markov models from event logs; (2)~a new technique (\emph{express analysis}) to calculate the mean of the  overall process execution time using these models; (3)~new methods (\emph{full analysis}) to calculate continuous and discrete distributions of the overall process execution time; (4)~analysis of the computational complexity of the full analysis technique for continuous and discrete cases, and (5) implementation and application of the proposed techniques to real-world event data to estimate the \emph{performance accuracy} of the discovery algorithms.

\section{Related Work}
\label{sec:rel_work}

Various approaches for the discovery of stochastic models (e.g., Markov chains or stochastic Petri nets), are presented in the literature~\cite{5f0e8dd04572478fb450eefe4211d69f,ROZINAT2009305,10.4108/icst.valuetools.2011.245715,10.1007/978-3-319-15895-2_41,7785764,10.1007/978-3-319-98648-7_10,10.1007/978-3-030-72693-5_20,10.1007/978-3-030-76983-3_16,9230330,opt_disc,gram_disc}. However, none of these approaches use mathematical tools offered by  discovered models to analyze process performance. \hl{Anastasiou et al.}~\cite{10.4108/icst.valuetools.2011.245715} presents a discovery method for inferring stochastic Petri nets with Erlang distributions from location tracking data. A technique for the discovery of $k$-order Markov chains from event logs is discussed in~\cite{10.1007/978-3-319-98648-7_10}. Hidden Markov models that capture both events and resources are discovered and analyzed in~\cite{10.1007/978-3-319-15895-2_41}. A method that discovers hidden semi-Markov models from event data is presented in~\cite{7785764}, this method predicts only the most frequent sequences of states given the initial steps of the process. Techniques that  discover stochastic Petri nets for their further simulation are presented in~\cite{5f0e8dd04572478fb450eefe4211d69f, ROZINAT2009305}. \hl{Burke et al.}~\cite{10.1007/978-3-030-72693-5_20,10.1007/978-3-030-76983-3_16} propose methods that discover stochastic Petri nets and assess quality of the discovered models by relating them to the event logs.  Besides stochastic process discovery, several methods for stochastic conformance checking were recently introduced in~\cite{10.1007/978-3-319-98648-7_10, Leemans2019,10.1007/978-3-031-16171-1_7,10.1007/978-3-030-49435-3_14,LEEMANS2023102197, leemans2023enjoy,9980707}.

Several performance analysis techniques proposed in the process mining context are based on the \emph{simulation} of discovered models~\cite{solti-predict-simulation,ROGGESOLTI20151,vandenabeele2022enhancing,ROZINAT2009305,CAMARGO2020113284}. A technique presented in~\cite{CAMARGO2020113284} is based on the discovery of BPMN (Business Process Model and Notation) models. These models are further enhanced by annotating activities with arbitrary time distribution functions inferred from event logs. This technique is focused on the adjustment of the discovery hyper-parameters to build accurate simulation models. A method to discover and simulate stochastic Petri nets with arbitrary distributed activity times is presented in~\cite{solti-predict-simulation,ROGGESOLTI20151}. A similar technique that considers normal distributions for activity execution times was proposed in~\cite{ROZINAT2009305}.  \hl{Vandenabeele et al.}~\cite{vandenabeele2022enhancing} focuses on the discovery of stochastic Petri nets from a set of traces with prefixes similar to the analyzed trace to predict the remaining time by simulating the discovered model. A review of different methods that analyze activity waiting and execution times based on data is presented in~\cite{wait_times}. One of such methods that also considers resources and priorities is proposed in~\cite{10.1007/978-3-031-34560-9_11}.

Methods for the \emph{remaining time prediction} were introduced in~\cite{VANDERAALST2011450,10.1007/978-3-642-13094-6_5,context-aware}.
These methods construct automata models from event logs annotating their states with remaining times and predict the remaining time for a trace by replaying it on a model.  \hl{Van der Aalst et al.}~\cite{VANDERAALST2011450} is additionally focused on the analysis of different levels of models' abstractions and discusses their impact on the remaining time prediction. Distinct automata models are constructed for different contexts (attributes in the event logs) in~\cite{context-aware}. These prediction methods are log-based and cannot be applied for \emph{what-if} analysis.

Performance analysis through \emph{visualizations}~\cite{DBLP:conf/bpm/DenisovFA18,10.1007/978-3-642-36285-9_23,10.1007/978-3-642-12186-9_15,10.1145/3202710.3203151} allows end users to get valuable insights into the performance of the process. Techniques for projecting performance information onto Petri nets, hierarchical models, and fuzzy models, discovered from event logs are implemented in~\cite{10.1007/978-3-642-36285-9_23}, \cite{10.1145/3202710.3203151}, and~\cite{10.1007/978-3-642-12186-9_15}, respectively. \hl{Denisov et al.}~\cite{DBLP:conf/bpm/DenisovFA18} presents a visualization technique where event log activities are visualized as lines with start and end points corresponding to the begin and end activity times, and the color corresponding to the duration of the activity. 

\emph{Closed-form solutions} for the performance analysis of models discovered from event logs are presented in~\cite{tree_performance,SENDEROVICH2019240,SENDEROVICH201896}. However, these solutions differ from the methods presented in this paper. \hl{Senderovich et al.}~\cite{SENDEROVICH201896} considers queuing networks and stochastic Petri nets with exponentially distributed execution times. An approach for deriving the most frequent sequences of states of hidden Markov models using Viterbi algorithm is applied in~\cite{SENDEROVICH2019240}.  \hl{Van Zelst et al.}~\cite{tree_performance} presents an algorithm that derives only time intervals for activities and sub-processes, and does not consider probabilities of events. In \hl{McClean et al.}~\cite{math11245001}, that considers semi-Markov models and execution times derived from event logs of smart home systems, distributions are represented in a form of mixture gamma models, the convolutions are only calculated for two pairs of transitions of artificial simulated data, where distributions were represented by three mixed gamma components. \hl{In\mbox{\cite{piulachs2024semimarkovmultistatemodelingapproaches}}, an analysis of a specific type of semi-Markov process based on Cox models is presented and applied to a real-world hospital event log. Various approaches to event history analysis in medicine, including Markov processes, Cox-based models, and tests of the Markovian property in these models, are discussed in in\mbox{\cite{doi:10.1191/0962280202SM276ra}} and \mbox{\cite{10.1093/biostatistics/kxaa030}} .}

In this paper, we propose \emph{general techniques} that calculate mean, as well as density probability functions for the \emph{overall process execution time} in the presence of arbitrary time distributions and  probabilities of events. Moreover, we apply our techniques to real-world event logs and demonstrate their applicability in practice.

Techniques that analyze \emph{performance characteristics of stochastic process models} outside the context of process mining are also presented in the literature. \hl{Kim et al.}~\cite{Kim2003EfficientCF} considers an approach for calculating $n$-th moments of process execution times for extended stochastic Petri nets using regular expressions. A relation between execution times in semi-Markov models and regular expressions was first discussed in~\cite{4505529}. A technique that calculates the execution time for block-structured stochastic time Petri nets is presented in \hl{Carnevali et al.}~\cite{10.1007/978-3-030-85172-9_5}.

In this work, we propose, implement and test \emph{most general approach} that constructs a probability density function of the overall process execution time, we also provide the exact closed-form solution for finding the mean process execution~time.
\section{Preliminaries}
\label{sec:prelim}

In this section, we give basic definitions that will be used to describe the proposed discovery approach.

Information systems  record their behavior in the form of event logs represented by sequences of events corresponding to (business) cases. Examples include event logs of web-servers, trouble-ticket, and application management systems. We formally define event logs as follows.

\begin{definition}[Event log]
Let $E$ be a finite non-empty set of events, $A$~be a set of activity names, $C$ be a finite set of cases, and $T$ be a set of timestamps.  
An \emph{event log} is a tuple $L=(E,f_a,f_c,f_t)$, where $f_a:E\rightarrow A$ is a function that assigns names to events, $f_c:E\rightarrow C$ is a surjective function that maps events to cases, and function $f_t:E\rightarrow T$ defines occurrence times for events.
\end{definition}
 
\autoref{tab:event-log} presents a fragment of an event log of an information system that processes applications. These can be applications for any kind of services, including government services, publishing, bug tracking, or any other requests managed by an information system.

\begin{table}[h!]
\footnotesize
\renewcommand\arraystretch{1.25}
%\vspace{-10pt}
\caption{\label{tab:event-log}A log of an application management system.}
\vspace{-10pt}
\begin{center}
\begin{tabular}{ l|c|c|c| } 
\cline{2-4}
                    & \textit{Activity name} & \textit{Case} & \textit{Timestamp} \\
\cline{2-4}
{\color{darkgray}$e_1$} & {\ttfamily"}Claim{\ttfamily"} & 1 & 2022-06-17 14:53:03 \\
\cline{2-4}
{\color{darkgray}$e_2$} & {\ttfamily"}Claim{\ttfamily"} & 2 & 2022-06-17 17:33:47  \\
\cline{2-4}
{\color{darkgray}$e_3$} & {\ttfamily"}Assign{\ttfamily"} & 1 & 2022-06-18 12:38:30  \\
\cline{2-4}
{\color{darkgray}$e_4$} & {\ttfamily"}Resolve{\ttfamily"} & 2 & 2022-06-19 09:46:03  \\
\cline{2-4}
{\color{darkgray}$e_5$} & {\ttfamily"}Close{\ttfamily"} & 2 & 2022-06-19 18:57:52  \\
\cline{2-4}
{\color{darkgray}$e_6$} & {\ttfamily"}Resolve{\ttfamily"} & 1 & 2022-06-20 10:37:52   \\
\cline{2-4}
{\color{darkgray}$e_7$} & {\ttfamily"}Close{\ttfamily"} & 1 & 2022-06-20 19:41:23  \\
\cline{2-4}
{\color{darkgray}$e_8$} & {\ttfamily"}Resolve{\ttfamily"} & 2 & 2022-06-21 18:14:51  \\
\cline{2-4}
{\color{darkgray}$e_9$} & {\ttfamily"}Assign{\ttfamily"} & 3 & 2022-06-21 22:56:05  \\
\cline{2-4}
{\color{darkgray}$e_{10}$} & {\ttfamily"}Resolve{\ttfamily"} & 3 & 2022-06-22 11:09:43  \\
\cline{2-4}
{\color{darkgray}$e_{11}$} & {\ttfamily"}Close{\ttfamily"} & 2 & 2022-06-22 17:49:46  \\
\cline{2-4}
{\color{darkgray}$e_{12}$} & {\ttfamily"}Close{\ttfamily"} & 3 & 2022-06-22 22:58:02  \\
\cline{2-4}
\end{tabular}
\end{center}
%\vspace{-20pt}
\end{table}
\normalsize

An application starts with a "Claim," which generates a {\em ticket}. The ticket is given a unique identifier called a case number, and is "Assigned" to a team member; then, the request is "Resolved"; and after that, the ticket is "Closed". Alternatively, the ticket can be reopened and the request is resolved once again.
Each row in \autoref{tab:event-log} represents an event from the finite set of events $E=\{e_1,e_2,\dots,e_{12}\}$ and each event is characterized by its name (\textit{Activity name} column), case or ticket identifier (\textit{Case} column), and the occurrence time (\textit{Timestamp} column). For example, event $e_1\in E$ is defined as follows: $f_a(e_1)=\text{{\ttfamily"}Claim{\ttfamily"}}$, $f_c(e_1)=1$, and $f_t(e_1)=\text{2022-06-17~14:53:03}$.

From an event log, we can develop {\em traces}.
We formally define this below, but loosely these are time-ordered sequences of events grouped by cases.

\begin{definition}[Trace of event log]
Let $L=(E,f_a,f_c,f_t)$ be an event log defined over a set of cases~$C$. A sequence of events $\delta=\langle e_1,e_2,\dots,e_n\rangle$, where $e_1,e_2,\dots,e_n\in E$, is called a \emph{trace} of~$L$ corresponding to case $c$ from $C$ iff it is the longest sequence, such that $\forall i,j$, $1\leq i\leq j\leq n$, it holds that $f_c(e_i)=f_c(e_j)=c$ and $f_t(e_i)\leq f_t(e_j)$.

\end{definition}

Note that if two events of the same case have equal timestamps, they can be ordered differently.

\begin{definition}[Trace representation of event log]
A \emph{set} of all traces of log $L$ corresponding to all its cases is a \emph{trace representation} of $L$, denoted as~$\overline{L}$.
\end{definition}

\begin{definition}[Activity/time traces]
\label{def:rep}
Let $L\nolinebreak=\nolinebreak(E,f_a,f_c,f_t)$ be an event log and $\delta\nolinebreak=\nolinebreak\langle e_1,e_2,\dots,e_n\rangle\in\overline{L}$ be one of its traces. Then we say that $\delta_A\nolinebreak=\nolinebreak\langle f_a(e_1),f_a(e_2),\dots,f_a(e_n)\rangle$ is an \emph{activity trace} and
$\delta_T=\langle f_t(e_1),f_t(e_2),\allowbreak \dots,f_t(e_n)\rangle$ is a \emph{time trace} corresponding to $\delta$.
\end{definition}
Activity and time traces have the same length as the original trace.
. 
 
The event log $L$ presented in \autoref{tab:event-log} can be transformed to its trace representation as follows: $\overline{L}=\{\delta^1,\delta^2,\delta^3\}$, where $\delta^1\nolinebreak=\nolinebreak\langle e_1,e_3,e_6,e_7\rangle$, $\delta^2=\langle e_2,e_4,e_5,e_8,e_{11}\rangle$, and $\delta^3\nolinebreak=\nolinebreak\langle e_9,e_{10},e_{12}\rangle$. The activity and time traces corresponding to $\delta^1$ are $\delta^1_A\nolinebreak=\nolinebreak\langle\text{{\ttfamily"}Claim{\ttfamily"}},\text{{\ttfamily"}Assign{\ttfamily"}},\text{{\ttfamily"}Resolve{\ttfamily"}},\text{{\ttfamily"}Close{\ttfamily"}}\rangle$ and $\delta^1_T=\langle t_1,t_2,t_3,t_4\rangle$, respectively, where, for example, $t_1\nolinebreak=\nolinebreak\scalebox{1.0}{2022-06-17 14:53:03}$ and $t_2\nolinebreak=\nolinebreak\scalebox{1.0}{2022-06-18 12:38:30}$.

Now we will introduce common notations used for different types of traces.
\begin{definition}[Subtrace, element and trace length]
Consider a trace of elements $\delta=\langle y_1,y_2,\dots,y_n\rangle$. A~\emph{subtrace} of $\delta$ is a sequence $\langle y_i,\dots, y_j\rangle$, $1\leq i\leq j\leq n$, denoted as $\delta(i,j)$. The $i^{th}$ element of trace $\delta$ is encoded as $\delta(i)$.  We denote the length of $\delta$ by $\mathit{len}(\delta)$.
\end{definition}

Performance analysis implies that the models discovered from event logs should contain both the control flow and the time perspectives. To capture these two perspectives, we will leverage Markov processes~\cite{Ross85}.

\begin{definition}[Finite-state Markov chain]
Consider a stochastic process $\{X_n, n=0,1,2,\dots\}$, where random variables $X_n$ take on a \emph{finite} number of possible values (\emph{states}). We say that $X_n=i$ when the stochastic process is in state $i$ at time $n$. We suppose that whenever the stochastic process is in state $i$, there is a fixed probability $p_{i,j}\in[0,1]$ that the next state will be $j$, such that $\sum_{j=1}^m p_{i,j}=1$, where $m$ is the number of states. 
By $\mathbf{P}=(p_{i,j})_{1\leq i,j\leq m}$ we denote the \emph{stochastic matrix}. We call the stochastic process a (finite-state) \emph{Markov chain} when 
$P\{X_{n+1}=j|X_n=i,X_{n-1}=i_{n-1},\dots X_1=i_1,X_0=i_0\}=P\{X_{n+1}=j|X_n=i\}=p_{i,j}$, \ie the choice of the next state depends only on the current state of the process.
The transition probability from state $i$ to state $j$ in $k>0$ steps is denoted as $p^k_{i,j}=P\{X_{k+l}=j|X_l=i\}$.
\end{definition}

Markov chains are uniquely defined by their stochastic matrices and can be represented as graphs where nodes correspond to states and direct arcs labeled by $p_{i,j}$ connect states $i$ and $j$  iff $p_{i,j}>0$.
\autoref{fig:simple_mc} presents a Markov chain with the stochastic matrix: 
$$
\label{eq:P}
\mathbf{P}= \begin{bmatrix}
  0 & 1 & 0\\ 
  0 & 0 & 1\\
  1 & 0 & 0
\end{bmatrix}.
$$

\begin{figure}[ht] 
\centering 
\begin{tikzpicture}[->,node distance=3cm]
\node[state] (q1) {1};
\node[state, right of=q1] (q2) {2};
\node[state, right of=q2] (q3) {3};
\draw (q1) edge[above] node{1} (q2)
      (q2) edge[above] node{1} (q3)
      (q3) edge[bend right, above] node{1} (q1);
\end{tikzpicture}
\caption{A Markov chain with three states.}
\label{fig:simple_mc}
\end{figure}

\begin{definition}[Irreducible Markov chain]
A Markov chain is called \emph{irreducible} iff every state  $i$ can be reached from any other state $j$, \ie there exists $k$, such that $p^k_{j,i} > 0$, or there is a path from $j$ to $i$ in its graph representation.
\end{definition}

The Markov chain in \autoref{fig:simple_mc} is irreducible, because all states are reachable from each other.

\begin{definition}[State period]
The period $d$ of a state $i$ is the greatest common divisor of the set $W_i=\{n:p^n_{i,i}>0\}$. A~state with period 1 is said to be \emph{aperiodic}.
\end{definition}

If an \emph{irreducible} Markov chain contains a state of period $d$, then all its states have period $d$, and $d$ is called the \emph{period} of the Markov chain~\cite{Ross85}.
All the states of the Markov chain in \autoref{fig:simple_mc} have period 3, because for any $i\in\{1,2,3\}$, $W_i=\{3,6,9\dots\}$, and so the Markov chain has period 3.

\begin{definition}[Aperiodic Markov chain]
A Markov chain $M$ is called \emph{aperiodic} iff all its states are aperiodic.
\end{definition}

\begin{figure}[ht] 
\vspace{-10pt}
\centering 
\begin{tikzpicture}[->,node distance=3cm]
\node[state] (q1) {1};
\node[state, right of=q1] (q2) {2};
\node[state, right of=q2] (q3) {3};
\draw (q1) edge[above] node{1} (q2)
      (q2) edge[above] node{0.5} (q3)
      (q2) edge[bend left, below] node{0.5} (q1)
      (q3) edge[bend right, above] node{1} (q1);
\end{tikzpicture}
\caption{An aperiodic Markov chain.}
\label{fig:aperiodic_mc}
\end{figure}

\autoref{fig:aperiodic_mc} presents an aperiodic Markov chain, as the greatest common divisor for $W_1=W_2=\{2,3,4,\dots\}$, as well as for $W_3=\{3,5,7,\dots\}$, is 1.

\begin{theorem}[Limiting probabilities~\cite{Ross85}]
\label{theor:lim}
For an irreducible Markov chain with period $d$, \emph{limiting probabilities} $\pi_j=\frac{1}{d}\lim\limits_{n\to\infty} p^{dn}_{i,j}$ exist for all states $j$, and are independent of initial states $i$. The $\pi_j$ are the unique non-negative solutions of $\pi_j=\sum_{i=1}^m \pi_i p_{i,j}$ such that $\sum_{j=1}^m \pi_j = 1$, where $m$ is the number of states.
\end{theorem}

The limiting probability is a \emph{long-run chance} that the process is in a particular state. For the Markov chain in \autoref{fig:aperiodic_mc}, the system of linear equations has the unique solution: $\pi_1=\sfrac{2}{5}$, $\pi_2=\sfrac{2}{5}$, and $\pi_3=\sfrac{1}{5}$.  That is, the  process visits state $1$ (or state $2$) twice as often as state~$3$.

In the \emph{business process management} (BPM) and \emph{process mining} (PM) disciplines, processes are usually considered as a ``set of interrelated or interacting activities which transforms inputs into outputs''~\cite{iso}. It is assumed that the process {\em starts} by taking inputs as parameters, then a sequence of steps is performed and outputs are produced and the process {\em stops}, unlike irreducible Markov process models. Such workflow models with specified start and end nodes were introduced and analyzed in~\cite{10.1007/3-540-47961-9_37,DBLP:journals/corr/abs-1812-07334}. To bring the stochastic modelling tools to bear we will define and analyze \emph{Markov process flows} that have a start and an end state, but such that the end connects to the start to create an irreducible Markov chain. 

\begin{definition}[Markov process flow]
A Markov chain is called a \emph{Markov process flow} iff it is irreducible and has a start $s$ and an end $e$ state, such that, $s\neq e$, $p_{e,s}=1$, \ie state $e$ has only one outgoing transition that leads to state $s$. Markov process flows will be denoted by 3-tuples $(\mathbf{P},s,e)$. 
\end{definition}

\begin{figure}[ht] 
\vspace{-10pt}
\centering
\begin{tikzpicture}[->,node distance=1.5cm]
\node[state] (q1) {s};
\node[state, right of=q1,draw=none, fill=white] (q2) {\dots};
\node[state, right of=q2] (q3) {i};
\node[state, right of=q3,draw=none, fill=white] (q4) {\dots};
\node[state, right of=q4] (q5) {e};
\draw (q1) edge[above] node{} (q2)
      (q2) edge[above] node{} (q3)
      (q3) edge[above] node{} (q4)
      (q4) edge[above] node{} (q5)
      (q5) edge[bend right, above] node{1} (q1);
\end{tikzpicture}
\caption{A Markov process flow.}
\label{fig:sm_flow}
\end{figure}

\autoref{fig:sm_flow} shows a Markov process flow. A Markov process flow is an irreducible Markov chain, and hence, every state $i$ is on a path from $s$ to $e$.

In real-world applications, the time spent in the process's states can vary. Moreover, the durations of different activities can follow different distribution laws~\cite{5f0e8dd04572478fb450eefe4211d69f}. To cover the time perspective of real-world processes, we will consider process models that assume that the time spent in a particular state is distributed according to an arbitrary distribution law. These are called semi-Markov processes and are formally defined below.

\begin{definition}[Semi-Markov process]
Consider a Markov chain with a stochastic matrix $\mathbf{P}=(p_{i,j})_{1\leq i,j\leq m}$. Let $\mathbf{F}=(f_{i,j}(t))_{1\leq i,j\leq m}$, $t\in\mathbb{R}$, be a \emph{waiting time} matrix and $f_{i,j}(t)$ be probability density functions (PDFs)\footnote{We assume that such probability density functions exist.} of the waiting time distribution when moving from state $i$ to state~$j$. A semi-Markov process is the Markov chain together with the probability density functions $f_{i,j}(t)$.
\end{definition}

Semi-Markov processes are also visualized as graphs and can be characterized as irreducible and aperiodic in respect to their stochastic matrices $\mathbf{P}$ just as for Markov processes.

If the waiting times are exponentially distributed, \hl{\emph{i.e.}, $f_{i,j}(t)$ are exponential density functions of the form $f_{i,j}(t) = \lambda_{i,j}e^{-\lambda_{i,j}t}$, where $\lambda_{i,j} > 0$ is the rate parameter,} then the semi-Markov process reduces to a continuous Markov process~\cite{Ross85}. \hl{As demonstrated in~\mbox{\autoref{sec:cont_time_dist}}, durations in event logs are not always exponentially distributed. Therefore, more general semi-Markov models are required.}

Semi-Markov processes are uniquely defined by their stochastic  matrix $\mathbf{P}$ and waiting time matrix $\mathbf{F}$, so we encode them as $(\mathbf{P},\mathbf{F})$ pairs. 
As with Markov processes we can extend semi-Markov processes to our flow model as follows.

\begin{definition}[Semi-Markov process flow]
A semi-Markov process $(\mathbf{P},\mathbf{F})$ is called \emph{semi-Markov process flow} with start $s$ and end $e$ states iff $\mathbf{P}$ encodes a Markov process flow with start and end states $s$ and $e$, respectively. 
\end{definition}

We will denote semi-Markov process flows as~4-tuples $(\mathbf{P},\mathbf{F},s,e)$.

For the analysis of overall process execution times we will use the notion of a convolution of continuous and discrete PDFs defined below. 

\begin{definition}[Convolution]
For two PDFs $f(t)$ and $g(t)$ of waiting times, $t\in\mathbb{R}$ (if $f(t)$ and $g(t)$ are continuous) and $t\in\mathbb{N}$ (if $f(t)$ and $g(t)$ are discrete), the PDF of the sum of the waiting times will be a \emph{convolution} $f(t)*g(t)$, such that: 
\[ f(t)*g(t)=\int\limits_{-\infty}^{+\infty}f(\tau)g(t-\tau)d\tau , \text{ in the continuous case,} \]

\[ f(t)*g(t)=\sum\limits_{\tau=-\infty}^{\tau=+\infty}f(\tau)g(t-\tau), \text{ } f(t), g(t) \text{ are discrete.}\]

By $\Conv\limits_{i=1}^m f(t)=f(t)*f(t)*\dots*f(t)$ we will denote a multiple $m$-times convolution of a function $f$ with itself\footnote{Note that it is perhaps more correct to denote a convolution as $[f * g](t)$, but the notation used above is common to much of the literature.}.
\end{definition}

\section{Process Discovery}
\label{sec:proces_discovery}
In this section, we describe how semi-Markov process flows can be discovered from event data.

\autoref{alg:stochastic:matrix:discovery} constructs stochastic and waiting time matrices by analyzing event logs \hl{and corresponds to the \emph{Discovery} and \emph{Enhancement} steps presented in~\mbox{\autoref{fig:schema}}.} 

%\vspace{-10pt}
\begin{algorithm}[h!]
\caption{SemiMarkovFlowDiscovery}
\label{alg:stochastic:matrix:discovery}
\KwIn{An event log $L$, an order $k$.}
\KwOut{A semi-Markov flow $(\mathbf{P},\mathbf{F},s,e)$.}
\BlankLine
$\overline{L} \leftarrow \textrm{ConstructTraceRepresentation}(L)$\;
$\mathbf{P} \leftarrow \mathbf{0};$ $\mathbf{F} \leftarrow \mathbf{nil}$ \tcp*[l]{initialize $\mathbf{P}$ and $\mathbf{F}$;}
\ForEach{$\delta\in\overline{L}$}
{
    $\delta_A \leftarrow \textrm{ConstructActivityTrace}(\delta)$;
    
    $\delta_T \leftarrow \textrm{ConstructTimeTrace}(\delta)$\;
    \tcc{Connect to start state $s$}
    $p_{s,\delta_A(1,1)} += 1$\;
    \BlankLine
    \tcc{Connect states, calculate waiting times}
    \For{$i\gets 1$ \KwTo $\mathit{len}(\delta)-1$}{
         $\Delta_T=\delta_T(i+1)-\delta_T(i)$\;
         \uIf{$i<k$}{
            $p_{\delta_A(1,i),\delta_A(1,i+1)} += 1$\;
            $\textrm{UpdatePDF}(f_{\delta_A(1,i),\delta_A(1,i+1)},\Delta_T)$\;
          }
          \Else{
            $p_{\delta_A(i-k+1,i),\delta_A(i-k+2,i+1)} += 1$\;
            $\textrm{UpdatePDF}(f_{\delta_A(i-k+1,i),\delta_A(i-k+2,i+1)},\Delta_T)$\;
        }
    }
    \BlankLine
    \tcc{Connect to end state $e$}
    \uIf{$k<\mathit{len}(\delta)$}{
        $p_{\delta_A(\mathit{len}(\delta)-k+1,\mathit{len}(\delta)),e} += 1$\;
    }
    \Else{
        $p_{\delta_A(1,\mathit{len}(\delta)),e} += 1$\;
    }
}
$p_{e,s}=1$\;
\tcc{The sum of elements of each row must equal 1}
 $\mathbf{P} \leftarrow \textrm{Normalize}(\mathbf{P})$\;
\tcc{If the waiting time is not defined set it to 0}
 \For{$i,j$}{
         \uIf{$f_{i,j}=nil$}{
$\textrm{UpdatePDF}(f_{i,j},0)$\;
          }
    }
\Return{$(\mathbf{P},\mathbf{F},s,e)$}\;
%}
\end{algorithm}

This algorithm takes integer $k>0$ as an additional input parameter that characterizes the ``memory'' of the process, \ie the last $k$ events define the current process state.

Similarly to the discovery technique of~\cite{van2010process}, \autoref{alg:stochastic:matrix:discovery} represents each state of the process as a sequence of $k$ event names and connects the states based on the frequencies of events. Another similar discovery method that also considers frequencies of events is Heuristic miner~\cite{tonbeta166}, however, it does not analyze performance characteristics of the process. 

First,  \autoref{alg:stochastic:matrix:discovery} constructs a trace representation of the input event log (line~1), initializes the stochastic matrix $\mathbf{P}$ with zero values, assuming that there are no transitions between any of the states, and labels elements of the waiting time matrix $\mathbf{F}$ as undefined (line~2).

Then, the algorithm traverses through each of the traces analyzing the order of activity names and timestamps and constructing activity and time traces (lines~4-5). Where less than $k$ events are considered (line~11), states are defined as sequences of all activity names seen so far, \ie $\delta_A(1,i)$, where $i$ is the current position, and connected to the next states $\delta_A(1,i+1)$. When two states are connected, the corresponding element of the stochastic matrix is increased by 1 (line~12). When more than $k$ activity names are considered, states $\delta_A(i-k+1,i)$, $\delta_A(i-k+2,i+1)$ ($i$ is the current position in the trace) that represented sequences of $k$ last activity names, are connected (line~15). As a result, the stochastic matrix contains weights of transitions (the number of the transitions' occurrences in the event log) and should be normalized to satisfy the property that the sum of elements of each row is~1~(line~28).

As the algorithm goes through the sequences of activity names, it also analyzes the corresponding time traces. Whenever a new connection between the states is added, the time difference between the occurrences of the event in the position $i+1$ and the event in the position $i$ is calculated (line~10). We assume that the waiting time for a transition is the difference in times of the current and the next event. After the waiting time is calculated, the corresponding PDF is updated taking into account this new time value (lines~13,16). \hl{In this pseudo-code, the $UpdatePDF$ function is abstract and can be implemented in various ways.} In practice, it is feasible to first collect all the values associated with the transition and then infer the PDF only once. There exist various approximation methods for PDFs~\cite{5f0e8dd04572478fb450eefe4211d69f,10.2307/2241953,Wilson72numericalmethods}. However, \hl{in this work,} we use performance analysis focused on the approximations of PDFs with either Gaussian Mixture Models (see \autoref{sec:gaussians} for details) or with histograms (\autoref{sec:disc_time_dist}), as this allows us to model arbitrary waiting times in continuous or discrete forms, respectively.
\hl{Note that the space complexity of the proposed algorithms is linear with respect to the number of states and transitions, as well as the number of GMM components or the size of the support.}

The start state $s$ and end state $e$ are connected to the states corresponding to the first (line~7) and the last events (lines~19--24) in the log. These connections are proportional to the frequencies of the first and the last events. The backward transition from $e$ to $s$ is added in line~26. \hl{In line 28, the probability matrix 
\mbox{$\mathbf{P}$} is normalized so that the sum of the elements in each row equals 1.} Note that the waiting time is not defined for the transitions from and to the start and end states, and in these cases, we set their waiting times to zero (lines~29--33). By the construction, each state lies on a path from $s$ to $e$, and thus this backward transition creates an irreducible semi-Markov process.

\hl{Excluding PDF approximations, which will be discussed later, the time complexity of Algorithm~\mbox{\ref{alg:stochastic:matrix:discovery}} (model discovery) is linear with respect to the total length of all traces, regardless of the parameter 
$k$. A detailed evaluation of how 
$k$ impacts the performance and accuracy of the overall time analysis is provided in \mbox{\autoref{sec:all-case-studies}}.}

\autoref{fig:flow_ex_k=1} visualizes a stochastic matrix $\mathbf{P}$ with designated states $s$ and $e$ discovered by \autoref{alg:stochastic:matrix:discovery} from the event log~$L$ (see~\autoref{sec:prelim}) when the parameter $k$ is set to~1.

\begin{figure} 
\centering 
\begin{tikzpicture}[->,scale=0.78, every node/.style={scale=0.78}, node distance=2.5cm]
\node[state] (q1) {s};
\node[state, right of=q1,label={$\langle\text{{\ttfamily"}Claim{\ttfamily"}}\rangle$}] (q2) {1};
\node[state, right of=q2,label={$\langle\text{{\ttfamily"}Resolve{\ttfamily"}}\rangle$}] (q3) {2};
\node[state, right of=q3,label={$\langle\text{{\ttfamily"}Close{\ttfamily"}}\rangle$}] (q4) {3};
\node[state, below=7mm of q2, label=below:{$\langle\text{{\ttfamily"}Assign{\ttfamily"}}\rangle$}] (q6) {4};
\node[state, right of=q4] (q5) {e};
\draw (q1) edge[above] node{0.66} (q2)
      (q2) edge[above] node{0.5} (q3)
      (q3) edge[above] node{1} (q4)
      (q4) edge[above] node{0.75} (q5)
      (q2) edge[right] node{0.5} (q6)
      (q1) edge[left] node[xshift=4,yshift=-5]{0.33} (q6)
      (q6) edge[right] node[xshift=-3,yshift=-5]{1} (q3)
      (q5) edge[bend right=45, above] node{1} (q1)
      (q4) edge[bend left=45, below] node{0.25} (q3);
\end{tikzpicture}
\caption{A Markov process flow discovered from event log $L$ with $k$ set to 1.}
\label{fig:flow_ex_k=1}
\end{figure}

Increasing the parameter $k$ will lead to process models in which the next state is defined by the last $k$ events in the trace. Such models are known as Markov processes  of order $k$~\cite{Ching2004HigherorderMC}.

\autoref{fig:flow_ex_k=2} presents a process discovered by \autoref{alg:stochastic:matrix:discovery} for $k$=2, when the maximum length of the sequence that defines the next state is 2.

\begin{figure} 
\centering 
\begin{tikzpicture}[->,scale=0.78, every node/.style={scale=0.78},node distance=2.75cm]
\node[state,label={[yshift=-55]$\langle\text{{\ttfamily"}Claim{\ttfamily"}}\rangle$}] (q1) {1};
\node[state, right of=q1,label={[label distance=0.4cm]$\langle\text{{\ttfamily"}Claim{\ttfamily"}},$},
label={[xshift=4]$\text{{\ttfamily"}Assign{\ttfamily"}}\rangle$}] (q2) {2};
\node[state, right of=q2,label={[label distance=0.4cm,xshift=-20,yshift=-60]$\langle\text{{\ttfamily"}Assign{\ttfamily"}},$},
label={[xshift=-14,yshift=-60]$\text{{\ttfamily"}Resolve{\ttfamily"}}\rangle$}] (q3) {3};
\node[state, right of=q3,label={[label distance=0.4cm, xshift=-28,yshift=0]$\langle\text{{\ttfamily"}Resolve{\ttfamily"}},$},
label={[xshift=-26,yshift=0]$\text{{\ttfamily"}Close{\ttfamily"}}\rangle$}] (q4) {4};

\node[state, above= 8mm of q1] (q6) {s};
\node[state,above= 8mm of q3, label={[xshift=-5,yshift=-2]$\langle\text{{\ttfamily"}Assign{\ttfamily"}}\rangle$}] (q7) {6};
\node[state, above= 8mm of q4] (q8) {e};
\node[state, below= 10mm of q4,label={[label distance=0.4cm,yshift=-66]$\langle\text{{\ttfamily"}Close{\ttfamily"}},$},
label={[xshift=9,yshift=-66]$\text{{\ttfamily"}Resolve{\ttfamily"}}\rangle$}] (q9) {7};
\node[state, below = 10mm of q3,label={[label distance=0.4cm,yshift=-66]$\langle\text{{\ttfamily"}Claim{\ttfamily"}},$},
label={[xshift=7,yshift=-66]$\text{{\ttfamily"}Resolve{\ttfamily"}}\rangle$}] (q10) {8};

\draw (q1) edge[above] node{0.5} (q2)
      (q2) edge[above] node{1} (q3)
      (q3) edge[above] node{1} (q4)
      %(q4) edge[above] node{0.75} (q5)
      (q6) edge[right] node{0.66} (q1)
      (q6) edge[above] node{0.33} (q7)
      (q7) edge[right] node{1} (q3)
      (q8) edge[bend right=45, above] node{1} (q6)
      %(q5) edge[left] node{1} (q8)
      (q9) edge[bend right=30, right] node{1} (q4)
      (q4) edge[left] node{0.25} (q9)
      (q1) edge[bend right=15, below] node{0.5} (q10)
      (q4) edge[right] node{0.75} (q8)
      (q10) edge[above] node{1} (q4);
\end{tikzpicture}
\caption{A Markov process flow discovered from $L$ with $k$ set to 2.}
\label{fig:flow_ex_k=2}
\end{figure}
\section{Performance Analysis}

This section presents two types of performance analysis. First, we present \emph{express analysis} for the calculation of the mean of the process execution time; this type of analysis also provides additional information on the contribution of each of the process states. Second, we present \emph{full analysis} that calculates a PDF for the overall process execution time.

\subsection{Express analysis. Deriving Mean of Process Execution Time}
\label{sec:express_analysis}
In this subsection, we propose new formulas to estimate the overall process execution time. These results provide analytical solutions that also assess the contribution of process states to the overall process execution time and are based on their limiting probabilities. These results can be further used in a \emph{what-if} analysis. 

First, we provide specific properties of limiting probabilities for Markov process flows that will be used later in this subsection.
Let $\mathbf{P}=(p_{i,j})$ be a stochastic matrix of a Markov process flow $(\mathbf{P},s,e)$. Then, by $\widehat{\mathbf{P}}=(\widehat{p}_{i,j})$ we will denote a matrix, such that, $\widehat{p}_{i,j}=0$, if $i=e, j=s$, and $\widehat{p}_{i,j}=p_{i,j}$, otherwise. This matrix assumes that there is no edge from $e$ to~$s$. By  $\widehat{p}^{\text{ }l}_{i,j}$ we denote a transition probability in $l$~steps from $i$ to $j$ in this Markov chain.

\begin{lemma}[Properties of limiting probabilities of Markov process flows]
\label{lemma}
 Let $(\mathbf{P},s,e)$~be a Markov process flow and $\pi_r$ be a limiting probability for state $r$. Then $\pi_r=\pi_s\sum_{l=0}^\infty \widehat{p}_{s,r}^{\text{ }l}$, \ie the limiting probability of state $r$ can be defined as the limiting probability of the start state multiplied by the sum of all transition probabilities from $s$ to $r$ without transiting through $(e,s)$ edge.  
 \begin{proof}
 See~\autoref{sec:appendix}.
 \end{proof}
\end{lemma}

\autoref{lemma} provides results that can be used for the calculation of the mean of the process execution time.

\begin{theorem}[Mean of process execution times]
\label{theor:mean}
Let $(\mathbf{P},\mathbf{F},s,e)$ be a semi-Markov process flow. Then, the mean time~$\mu$ of hitting time for state $e$ when the process has started in state $s$ (the mean of the overall execution time) can be calculated as follows: 
\begin{equation}
\label{eqn:mean}
\mu=\frac{1}{\pi_s}\sum\limits_{{\substack{i=1, \\ i\neq e}}}^{m}\pi_i\mu_i \,,
\end{equation}
where $\pi_s$ is the limiting probability for the start state $s$, $\pi_i$ and $\mu_i$ are the limiting probability and the mean hitting time, respectively, for the state $i$, and $i$ iterates over all $m$ states except end state $e$.
\begin{proof}
See~\autoref{sec:appendix}.
\end{proof}
\end{theorem}

\autoref{theor:mean} provides results that can be used in practice to estimate the overall process execution time based on the time distributions for each of the states. This method does not rely on any particular PDF approximation, allowing estimation of the mean time using simple formulas for a finite number of observations. 

Let $[\Delta_{T_1},\Delta_{T_2},\dots,\Delta_{T_n}]$ be a multiset of time durations when moving from state $i$ to state $j$, extracted from the event log (\autoref{alg:stochastic:matrix:discovery}). The $\mathit{UpdatePDF}$ function can be implemented to recalculate the mean values $\mu_i=\sum\limits_{j=1}^m p_{i,j}(\sfrac{\sum\limits_{r=1}^n \Delta_{T_r}}{n})$ each time new durations are added.
After that, limiting probabilities are retrieved as a solution of linear equations (\autoref{theor:lim}) and a mean value $\mu=\frac{1}{\pi_s}\sum_{{\substack{i=1, i\neq e}}}^{m}\pi_i\mu_i$ of the overall time is calculated (\autoref{theor:mean}).

This general approach can be applied when PDFs are known or inferred using an approximation algorithm.
It can also be applied for a \emph{what-if} analysis, allowing one to estimate the impact of each of the states on the overall process execution time.

Consider an example of event log (\autoref{tab:event-log}) and a corresponding Markov process flow discovered from this event log with $k$ set to 1 (\autoref{fig:flow_ex_k=1}). The mean waiting times for the states extracted from the event log are: $\mu_1$=1d\,:\,6h:58m\,:\,52s, $\mu_2$=13h\,:\,24m\,:\,39s, $\mu_3$=11h\,:\,49m\,:\,15s,
$\mu_4$=1d\,:\,5h\,:\,6m\,:\,30s, $\mu_s=\mu_e=0$.  The limiting probabilities are the following: $\pi_1=\sfrac{1}{9}$, $\pi_2=\sfrac{2}{9}$, $\pi_3=\sfrac{2}{9}$,
$\pi_4=\sfrac{1}{9}$, and $\pi_s=\pi_e=\sfrac{1}{6}$. According to \autoref{eqn:mean}, the mean of the overall process execution time is 3d\,:\,1h\,:\,42m\,:\,5s. 

Now a \emph{what-if} analysis can be run, and the process can be optimized in two ways: (1) reducing the waiting time for some of the states (reducing $\mu_i$ values); (2) reorganizing the structure of the process (changing $\pi_i$ values). 

Firstly, to optimize the process we may try to reduce waiting times for the states with the highest impact (highest limiting probabilities), these are states $2$ and $3$, or highest waiting times, these are states $1$ and $4$. Reducing waiting times for states $1$ and $4$ means we reduce the time before the event labeled by the activity name $\text{{\ttfamily"}Resolve{\ttfamily"}}$ occurs. This can be achieved by assigning more resources (employees) who resolve the requests. Suppose that the waiting times for states $1$ and $4$ were halved, and the new mean times are $\mu'_1$=15h\,:\,29m\,:\,26s and $\mu'_4$=14h\,:\,33m\,:\,15s, respectively.
Then, according to \autoref{eqn:mean}, the new mean of the overall process execution time is 2d\,:\,5h\,:\,40m\,:\,19s, which is considerable improvement comparing to the original time 3d\,:\,1h\,:\,42m\,:\,5s.

Another way to optimize the process is to modify the limiting probabilities changing the stochastic matrix, and hence, the structure of the process. A Markov process flow (\autoref{fig:flow_ex_k=1}) assumes that in half of the cases a ticket that was claimed by an employee can be reassigned (or approved) later. This can lead to potential delays. It may be feasible to allow most of the employees to start working on applications they have already requested without further approval. An optimized Markov process flow in which only 10\% of ticket claims are to be reassigned (approved) is presented in~\autoref{fig:flow_ex_k=1_opt}. In this case the limiting probabilities will be:  $\pi'_1=\sfrac{10}{86}$, $\pi'_2=\sfrac{20}{86}$, $\pi'_3=\sfrac{20}{86}$,
$\pi'_4=\sfrac{6}{86}$, and $\pi'_s=\pi'_e=\sfrac{15}{86}$, and, according to~\autoref{eqn:mean}, the mean of the overall process execution time is:    2d\,:\,17h\,:\,56m\,:\,23s, which is less than the mean of the overall time (3d\,:\,1h\,:\,42m\,:\,5s) for the original process.            
            
\begin{figure}[ht] 
\centering 
\begin{tikzpicture}[->,scale=0.78, every node/.style={scale=0.78},node distance=2.5cm]
\node[state] (q1) {s};
\node[state, right of=q1,label={$\langle\text{{\ttfamily"}Claim{\ttfamily"}}\rangle$}] (q2) {$1'$};
\node[state, right of=q2,label={$\langle\text{{\ttfamily"}Resolve{\ttfamily"}}\rangle$}] (q3) {$2'$};
\node[state, right of=q3,label={$\langle\text{{\ttfamily"}Close{\ttfamily"}}\rangle$}] (q4) {$3'$};
\node[state, below=7mm of q2, label=below:{$\langle\text{{\ttfamily"}Assign{\ttfamily"}}\rangle$}] (q6) {$4'$};
\node[state, right of=q4] (q5) {e};
\draw (q1) edge[above] node{0.66} (q2)
      (q2) edge[above] node{0.9} (q3)
      (q3) edge[above] node{1} (q4)
      (q4) edge[above] node{0.75} (q5)
      (q2) edge[right] node{0.1} (q6)
      (q1) edge[left] node[xshift=4,yshift=-5]{0.33} (q6)
      (q6) edge[right] node[xshift=-3,yshift=-5]{1} (q3)
      (q5) edge[bend right=45, above] node{1} (q1)
      (q4) edge[bend left=45, below] node{0.25} (q3);
\end{tikzpicture}
\caption{An optimized Markov process flow corresponding to the original process model presented in~\autoref{fig:flow_ex_k=1}. In contrast to~\autoref{fig:flow_ex_k=1}, transitions between states $1'$ and $4'$, $1'$ and $2'$, have 0.1, 0.9 probabilities, respectively.}
\label{fig:flow_ex_k=1_opt}
\end{figure}

Although the model discovered by the proposed algorithm allows for more behavior in terms of sequences of events and their execution times, the mean time of one process run for the discovered model and the mean trace time for the corresponding event log are the same.

\begin{theorem}[Log and model mean times]
\label{theor:mean_eq}
Let $L$ be an event log and $M=(\mathbf{P},\mathbf{F},s,e)$ be a semi-Markov process flow discovered from this log by \autoref{alg:stochastic:matrix:discovery}, with an arbitrary order $k$ set as a parameter. Then, the mean process execution times for the log and the model are the same, \ie $\mu_M=\mu_L$, where $\mu_M$ is the mean $e$ hitting time and $\mu_L$ is the mean duration of a trace for the event log $L$.  
\begin{proof}
See~\autoref{sec:appendix}.
\end{proof}
\end{theorem}

Note that the mean time for the event log (\autoref{tab:event-log}) is the same as for the model discovered from this event log (\autoref{fig:flow_ex_k=1}), and it equals 3d\,:\,1h\,:\,42m\,:\,5s.

\subsection{Full Analysis Of The Process Execution Time}
\label{sec:full_analysis}
This section presents a method that calculates a PDF for the overall process execution time using a so-called \emph{reduction} technique. This reduction technique is based on the algorithm that builds regular expressions from finite state automata models~\cite{hopcroft2001introduction}.  Conversion of semi-Markov models to regular expressions using this type of algorithm was first discussed in~\cite{4505529}. However, it was not implemented and was not applied to real-world data. In this work, we provide and test a complete reduction algorithm for the construction of PDFs for the overall execution time of semi-Markov process flows.

Consider a fragment of a semi-Markov process flow that is represented by a graph, where nodes correspond to the process states and edges are labeled by transition probabilities and PDFs of waiting times (\autoref{fig:mc_reduction_a}).

\begin{figure}
\centering
\begin{tikzpicture}[->,node distance=5.0cm]
\node[state,label={}] (q2) {1};
\node[state, below right=25.5mm of q2,label={}] (q3) {2};
\node[state, above right=25.5mm of q3,label={}] (q4) {4};
\node[state, below left=25.5mm of q3, label={}] (q6) {3};
\node[state, below right=25.5mm of q3, label={}] (q5) {5};
\draw (q2) edge[above] node[xshift=31, yshift=-2]{$(p_{1,2},f_{1,2}(t))$} (q3)
      (q3) edge[above] node[xshift=45, yshift=-5]{$(p_{2,4},f_{2,4}(t))$} (q4)
      (q6) edge[above] node[xshift=35, yshift=-15]{$(p_{3,2},f_{3,2}(t))$} (q3)
      (q3) edge[above] node[xshift=40, yshift=-12]{$(p_{2,5},f_{2,5}(t))$} (q5)
      (q3) edge[loop left] node[xshift=0, yshift=0]{$(p_{2,2},f_{2,2}(t))$} (q3);
\end{tikzpicture}
\vspace{11pt}
\caption{A fragment of a semi-Markov process flow.}
\label{fig:mc_reduction_a}
\end{figure}

\begin{figure}
\centering
\begin{tikzpicture}[->,node distance=1.0cm]
\node[state, label={}] (q2) {1};
\node[state, right=67mm of q2,label={}] (q4) {4};
\node[state, below=35mm of q2, label={}] (q6) {3};
\node[state, right=67mm of q6, label={}] (q5) {5};
\draw (q2) edge[above] node[xshift=-15, yshift=0]{$(p_{1,2}\frac{p_{2,4}}{p_{2,4}+p_{2,5}},f_{1,2}(t)\!*\!\overset{\scaleobj{0.9}{\leftturn}}{f_2}(t)\!*\!f_{2,4}(t))$} (q4)
(q6) edge[above] node[xshift=-25, yshift=-25]{$(p_{3,2}\frac{p_{2,5}}{p_{2,4}+p_{2,5}},f_{3,2}(t)\!*\!\overset{\scaleobj{0.9}{\leftturn}}{f_2}(t)\!*\!f_{2,5}(t))$} (q5)
(q2) edge[above] node[xshift=0, yshift=-60]{($p_{3,2}\frac{p_{2,4}}{p_{2,4}+p_{2,5}},f_{3,2}(t)\!*\!\overset{\scaleobj{0.9}{\leftturn}}{f_2}(t)\!*\!f_{2,4}(t))$} (q5)
(q6) edge[above] node[xshift=0, yshift=37]{$(p_{1,2}\frac{p_{2,5}}{p_{2,4}+p_{2,5}},f_{1,2}(t)\!*\!\overset{\scaleobj{0.9}{\leftturn}}{f_2}(t)\!*\!f_{2,5}(t))$} (q4);
\end{tikzpicture}
\caption{The semi-Markov process flow after reduction of state $2$.}
\label{fig:mc_reduction_b}
\end{figure}

The reduction algorithm removes all the states (except start and end states) adding new transitions corresponding to the removed paths. 
\autoref{fig:mc_reduction_a} presents a fragment of a semi-Markov process flow before the reduction. The corresponding fragment after the elimination of state $2$ is shown in~\autoref{fig:mc_reduction_b}. Once state $2$ is removed, all the states that were connected via this state, will be connected directly. Corresponding transitions will be labeled with new probabilities and PDFs. For example, the new transition that links states~$1$ and $4$ is labeled by $\sfrac{p_{1,2}p_{2,4}}{(p_{2,4}+p_{2,5})}$ -- the probability of transitioning from state $1$ to state $4$ via state $2$ in the original semi-Markov model. The new PDF for the transition time from $1$ to $4$ via state $2$ is defined as $f_{1,2}(t)\!*\!\overset{\scaleobj{0.9}{\leftturn}}{f_2}(t)\!*\!f_{2,4}(t))$, where $\overset{\scaleobj{0.9}{\leftturn}}{f_2}(t)=(1-p_{2,2})\left(f^0(t)+\sum_{\boldsymbol{s}=1}^\infty p_{2,2}^{\boldsymbol{s}}\Conv_{r=1}^{\boldsymbol{s}}f_{2,2}(t)\right)$ is a PDF for the looping time in state $2$, and 
\[ f^0(t)=\begin{cases}
    1, & \text{if $t=0$}.\\
    0, & \text{otherwise}.
  \end{cases}.
  \] 
Note that, states $1$ and $4$ are not necessarily distinct and the elimination of state~$2$ can lead to self-loops.

The case of merging multiple transitions leading to one state should be included to the general case, but for the sake of clarity, we consider this case separately. \autoref{fig:mc_reduction_mult_a} presents an initial fragment of a Markov process flow. The result of state~$2$ reduction, such that transitions between nodes $1$ and $3$ are merged is presented in \autoref{fig:mc_reduction_mult_b}. The probability for the new transition between states $1$ and $3$ will be defined as $p_{1,2}+p_{1,3}$ and the corresponding PDF will be a weighted sum of PDFs of waiting times for moving from state $1$ to $3$ directly and via state $2$. 

\begin{figure}[h!]
\label{fig:mc_reduction_mult}
\centering 
\begin{tikzpicture}[->]
\node[state,label={}] (q2) {1};
\node[state, right=30mm of q2,label={}] (q3) {2};
\node[state, right=30mm of q3,label={}] (q4) {3};
\draw (q2) edge[above] node[xshift=0, yshift=-2]{$(p_{1,2},f_{1,2}(t))$} (q3)
      (q3) edge[above] node[xshift=0, yshift=-2]{$(p_{2,3},f_{2,3}(t))$} (q4)
      (q2) edge[above, bend left=60] node[xshift=0, yshift=-2]{$(p_{1,3},f_{1,3}(t))$} (q4)
      (q3) edge[loop above] node[xshift=0, yshift=-2]{$(p_{2,2},f_{2,2}(t))$} (q3);
\end{tikzpicture}
\caption{A fragment of a Markov process flow with an additional transition.}
\label{fig:mc_reduction_mult_a}
\end{figure}

\begin{figure}[h!]
%\vspace{25pt}
\centering
\begin{tikzpicture}[->,node distance=5cm]
\node[state,label={}] (q2) {1};
\node[state, right=67mm of q2,label={}] (q4) {3};
\draw (q2) edge[above] node[xshift=0, yshift=15]{$
      (p_{1,2}+p_{1,3}$,
      $\frac{p_{1,2}}{p_{1,2}+p_{1,3}}f_{1,2}(t)\!*\!\overset{\scaleobj{0.9}{\leftturn}}{f_2}(t)\!*\!f_{2,3}(t)+\frac{p_{1,3}}{p_{1,2}+p_{1,3}}f_{1,3}(t))$} (q4);
\end{tikzpicture}
\caption{The Markov process flow without state $2$.}
\label{fig:mc_reduction_mult_b}
\end{figure}

This approach allows us to reduce the original Markov process flow iteratively so that only states $s$ and $e$  remain. The transition between these two states will be marked by a pair $(1, f_{\mathit{total}}(t))$, where $f_{\mathit{total}}(t)$ is the PDF for $e$ hitting time.

The representation of PDFs plays a crucial role in the analysis and inference of the overall process execution time, it also affects the computational complexity of the reduction algorithm. 
The next subsections discuss the approaches used in this paper for representing and manipulating PDFs of waiting times in the form of GMMs and discrete time distributions. Additionally, the computational complexity of each approach is provided.

In practice, the maximum number of self-loop repetitions is defined by a \textbf{loop threshold}, which sets the minimal probability for repeating a self-loop during the state reduction. For example, if the probability of repeating a self-loop in state~$2$ three times, $p^3_{2,2}$, is less than this threshold, the loop will not be repeated more than twice.

\subsubsection{Continuous Time Distributions}
\label{sec:gaussians}
As will be shown later, waiting times in real-world event data can follow a whole a range of distributions, including multi-modal distributions. Gaussian Mixture Models (GMMs) are ideal to encompass multi-modal data.
The PDFs of GMMs are of the form: $f(t) = \sum_{i=1}^\ell q_ig_i(t)\text{, where }\sum_{i=1}^\ell q_i=1\text{, and }$
$$ g_i(t)=\frac{1}{\sigma_i\sqrt{2\pi}}e^{-\frac{1}{2}\left(\frac{t-\mu_i}{\sigma_i}\right)^2},$$
is a PDF of a Gaussian distribution with mean $\mu_i$ and variance~$\sigma^2_i$. 

A key advantage of GMMs is that the convolution of two Gaussians $g(t)$ and $h(t)$ is a Gaussian $g(t)*h(t)$, such that $\mu_{g*h}=\mu_g+\mu_h$ and $\sigma^2_{g*h}=\sigma^2_g+\sigma^2_h$, \ie the mean and variance are sums of means and variances, respectively. This simplifies the calculation of convolution for two  GMMs, \eg according to the distributive property of convolution: $f(t)=(p_1g_1(t)+p_2g_2(t))*(p_3h_1(t)+p_4h_2(t))=p_1p_3(g_1(t)*h_1(t))+p_2p_3(g_2(t)*h_1(t))+p_1p_4(g_1(t)*h_2(t))+p_2p_4(g_2(t)*h_2(t))$.

 When the number of Gaussians is unbounded, the reduction algorithm's complexity can be estimated as $O(n^3\cdot\ell_0^{(\boldsymbol{s}+2)^n})$, where $n$ is the number of states, $\boldsymbol{s}$ is the maximum number of self-loop repetitions performed, and $\ell_0$ is the maximum number of Gaussians in each of the initial transitions. This is because each reduction considers GMMs associated with the input and output transitions, as well as up to $\boldsymbol{s}$ times associated with the self-loop transitions, and after each convolution the number of Gaussians squares. This estimation relates the proposed algorithm to the original algorithm that builds regular expressions from finite automata models, which has a time complexity of $O(n^3\cdot4^n)$~\cite{hopcroft2001introduction}.  

To lower the double exponential time complexity, we prune the number of Gaussians in GMMs using  a  \textbf{weight threshold}. After each state reduction, all the Gaussians in any new GMM with weight coefficients less than this threshold can be merged into one Gaussian with a mean and variance corresponding to the mean and variance of the weighted sum of the merged Gaussians. 

This optimization allows us to compute an upper bound for the time computational complexity of the reduction algorithm to be $O(n^3\cdot\ell^2 +n\cdot\boldsymbol{s}\cdot\ell^2)$, where $n$ is the number of states in the model, $\ell$ is the maximum number of Gaussians in a GMM, and $\boldsymbol{s}$ is the maximum number of self-loop repetitions. Moreover, assuming that $d$ is the state degree (the greater value between the number of input transitions and the number of output transitions), the computational complexity can be represented as $O(n\cdot d^2\cdot\ell^2 +n\cdot\boldsymbol{s}\cdot\ell^2)$. 

As the time distributions are represented only by mean and variance values and the number of Gaussian distributions is bounded, this approach allows us to perform computations for large-scale event data.
\hl{In~\mbox{\autoref{sec:case_studies}}, we demonstrate its practical impact on both accuracy and performance.}

There is one minor caveat --- Gaussian distributions have infinite support and as waiting times are non-negative we need, at some point, to truncate this representation. We discuss our approach to this truncation in our case study, described in~\autoref{sec:case_studies}.

\subsubsection{Discrete Time Distributions}
\label{sec:disc_time_dist}

Discrete PDFs are represented as probability mass functions $f(t)$, such that $f(t_i)=p_i$ if and only if $p_i$ is the probability of waiting time $t_i$. We will consider discrete time distributions with finite support that can be inferred from  event data in the form of histograms. Similarly to GMMs, the computational complexity of one convolution is quadratic\footnote{Long convolutions can be sped up using Fourier methods, but the direct approach is implemented in the libraries we use here, and this is faster unless the functions are quite long.} in respect to the length of the distributions but the length of the resulting probability mass function is the sum of the length of the input functions. Therefore the overall computational complexity can be calculated as $O(\sum\limits_{i=1}^n{(d^2+\boldsymbol{s})((\boldsymbol{s}+2)^{i-1}a_0)^2})$, which can be represented as $O(a_0^2(d^2+\boldsymbol{s})\sum\limits_{i=1}^n{(\boldsymbol{s}+2)^{2i-2}})=O(a_0^2(d^2+\boldsymbol{s})(\boldsymbol{s}+2)^{2n})$, where $n$ is the number of states, $a_0$ is the maximum size of support of the initial time distributions, $\boldsymbol{s}$ is the maximum number of self-loop repetitions, and $d$ is the maximum state degree. The value $(d^2+\boldsymbol{s})$ represents the maximum number of convolutions that need to be calculated when the state is reduced, $(\boldsymbol{s}+2)^{i-1}a_0$ indicates the maximum size of support of the time distributions obtained with this node reduction. Each new reduction increases the length of support by a factor of $(\boldsymbol{s} + 2)$ and requires a squared number of operations.   
Although the computational complexity is exponential, it manifests only for high-density graphs, when reduction of a state affects all other transitions, as all other states were connected through the removed state. 

\hl{Another way to represent the computational complexity is by incorporating the maximum size of the support of the time distributions into the formula. Suppose $a_m$ is the maximum size of the support; the computational complexity can then be expressed as $O(n\cdot a^2_m\cdot (d^2+\boldsymbol{s}))$. This aligns with the results presented in \mbox{\autoref{sec:case_studies}}, where we also provide the size of the support for the distributions obtained during the calculations.}
We also demonstrate that the approach based on discrete convolutions can outperform the bounded GMM-based approach for process models discovered from real-world event data, when the maximum or average support of the discrete time distributions is relatively small. Thus this approach might be preferred if measured times are highly discretised before we obtain them, but is not suitable when the continuously varying times are measured directly, or when the process model is densely connected. 

\label{sec:discrete}
\section{Case Studies}
\label{sec:all-case-studies}
In this section, we apply performance analysis techniques to real-world event data. The case study addresses practical components of the algorithm, such as the choice of the model order
$k$ and the representation of PDFs.

Firstly, we leverage the express analysis technique to estimate the process mean time and run a what-if analysis (\autoref{sec:express_case_study}).
We then explore how different types of time distributions in event data can be approximated (\autoref{sec:cont_time_dist}), and after that, we apply the  full performance analysis techniques to real-world event data using continuous and discrete convolutions, and analyze their accuracy (\autoref{sec:case_studies}). 

The proposed methods were implemented in Python using the PM4PY library~\cite{pm4py} and are available at: \\\url{https://github.com/akalenkova/TimeDistributions}.

\subsection{Applying Express Analysis to Real-World Data}
\label{sec:express_case_study}

\begin{figure*}[ht] 
\vspace{-20pt}
\centering % centers the figure
\begin{tikzpicture}[->,node distance=3.25cm]
\node[state] (q1) {s};
\node[state, right of=q1,label={[yshift=-60]$\langle\text{{\ttfamily"}Accepted{\ttfamily"}}\rangle$}] (q2) {1};
\node[state, right of=q2,label={[yshift=-1, xshift=13]$\langle\text{{\ttfamily"}Queued{\ttfamily"}}\rangle$}] (q3) {2};
\node[state, right of=q3,label={[yshift=0, xshift=35]$\langle\text{{\ttfamily"}Completed{\ttfamily"}}\rangle$}] (q4) {3};
\node[state, above=9mm of q2, label={[yshift=0, xshift=-15]$\langle\text{{\ttfamily"}Unmatched{\ttfamily"}}\rangle$}] (q6) {4};
\node[state, right of=q4] (q5) {e};
\draw (q1) edge[above] node{0.8465} (q2)
      (q2) edge[above, bend left=40] node{0.2368} (q3)
      (q3) edge[above] node{0.0032} (q4)
      (q4) edge[above] node{0.5442} (q5)
      (q5) edge[bend right=95, above] node{1} (q1)
      (q4) edge[bend left=48, above] node[yshift=3]{0.0081} (q3)
      (q1) edge[bend right=75, below, in=280] node{0.0005} (q4)
      (q1) edge[bend right=75, below, out=300, in=280] node{0.1530} (q3)
      (q2) edge [out=230,in=200,looseness=8] node[xshift=25,yshift=-6] {0.5615}(q2)
      (q2) edge[bend right=65, below, out=330] node[xshift=30,yshift=-2]{0.2015} (q4)
      (q2) edge[bend right=65, below, out=320] node[xshift=20, yshift=-2]{0.0002} (q5)
      (q3) to [out=230,in=200,looseness=8] node[xshift=-18,yshift=7] {0.0673}(q3)
      (q3) edge[above] node{0.9295} (q2)
      (q4) edge [out=330,in=300,looseness=8] node[xshift=0,yshift=-8] {0.4140}(q4)
      (q4) edge[bend right=53] node[yshift=10, xshift=5]{0.0004} (q6)
      (q4) edge[bend right=67] node[yshift=8]{0.0333} (q2)
      (q6) edge[left] node{1} (q2);
\end{tikzpicture}
\caption{A Markov process flow (where $s$ is a start state and $e$ is an end state) discovered from the log of an incident management system with k=1, \ie each state corresponds to a preceding sequence of activities of length 1.}
\label{fig:real_flow}
\end{figure*}

In this subsection, we consider event data from real-world incident management system's event log (\emph{BPI'13 incidents}~\cite{WardSteeman2013}) as a running example.
An incident management system of Volvo IT Belgium recorded the history of the production incident processes in the form of an event log~\cite{WardSteeman2013}. This event log contains 7,554 traces, 65,533 events, and 4 event names ("Accepted", "Queued", "Completed", or "Unmatched"). Each incident record was processed by a management team and the results were recorded in the log.
\autoref{fig:real_flow} shows a graphical representation of the Markov process flow discovered by \autoref{alg:stochastic:matrix:discovery} with the order $k$ set to~1.

Most commonly, once the process has started, the incident description is accepted, and then it is either accepted again
(for example, by a different employee) or the management procedure is completed. Alternatively, if there are not enough resources to process the request, it is queued. In rare cases, when the status of the request is unknown, the incident description is labeled as "Unmatched" and is considered again.

Once the model is discovered, a \emph{what-if} analysis can be applied.
By \autoref{theor:mean}, the mean time of the process is $\mu=\sfrac{1}{\pi_s}\sum_{{\substack{i=1,i\neq e}}}^{m}\pi_i\mu_i$, where $\pi_i$ and $\mu_i$ are the limiting probabilities and mean waiting times, respectively. This process is represented by the limiting probabilities and times: $[0.09367,\:\allowbreak 0.49748,\:\allowbreak 0.14316,\:0.17196,\: 0.00006,\: 0.09367]$, $[\text{0d 0h 0m 0s},\:\text{1d 5h 8m 9s},\:\text{1d 2h 9m 35s},\:\allowbreak\text{2d 3h 51m 27s},\allowbreak\text{0d 0h 11m 10s},  \text{0d 0h 0m 0s}]$ retrieved for the vector of process states: $[s,\langle\text{{\ttfamily"}Accepted{\ttfamily"}}\rangle,\langle\text{{\ttfamily"}Queued{\ttfamily"}}\rangle,\langle\text{{\ttfamily"}Completed{\ttfamily"}}\rangle,\allowbreak\langle\text{{\ttfamily"}Unmatched{\ttfamily"}}\rangle,\allowbreak e]$.

As can be calculated using $\mu=\sfrac{1}{\pi_s}\sum_{{\substack{i=1,i\neq e}}}^{m}\pi_i\mu_i$, the mean process execution time is $\text{12d 1h 54m 15s}$. The state with the largest value of $\pi_i\mu_i$ has the greatest impact on the overall performance of the process. In this case, this state is $\langle\text{{\ttfamily"}Accepted{\ttfamily"}}\rangle$. Although the state with the greatest mean time is $\langle\text{{\ttfamily"}Completed{\ttfamily"}}\rangle$, the state $\langle\text{{\ttfamily"}Accepted{\ttfamily"}}\rangle$ is the most time-consuming  because of its limiting probability \ie it occurs more often than other states.  The waiting time in the state $\langle\text{{\ttfamily"}Accepted{\ttfamily"}}\rangle$ corresponds to the actual incident management; if more staff were assigned to this process step,  it would reduce the overall mean time. If the mean time for the $\langle\text{{\ttfamily"}Accepted{\ttfamily"}}\rangle$ state were decreased by a factor of two, the average time of the entire process (calculated using the formula for the mean value) would be $\text{8d 20h 32m 19s}$.

Another way to improve the performance characteristics of the process is to reorganize its structure and adjust the transition probabilities. This will change the limiting probabilities, and hence, the overall execution time.

\subsection{Applying Full Analysis to Real-World Event Data}
In this subsection, we apply the full analysis technique to real-world event data. First, we propose methods to derive time distributions from event logs, \hl{we also demonstrate by a real-world example that durations in event logs can be far from exponentially distributed} (\autoref{sec:cont_time_dist}). Then, we apply the reduction technique to these event logs and assess the performance accuracy of the discovered models (\autoref{sec:case_studies}).

\label{sec:time_dist}
\subsubsection{Deriving Time Distributions}
\label{sec:cont_time_dist}
In this subsection, we demonstrate how PDFs of waiting times between the states can be derived from event data. As discussed in~\autoref{sec:full_analysis}, PDFs will be represented in continuous and discrete forms. 

\begin{figure*}[h!]
\vspace{-10pt}
\begin{subfigure}{.5\textwidth}
\centering\captionsetup{width=.9\linewidth}
\includegraphics[width=1.0\textwidth]{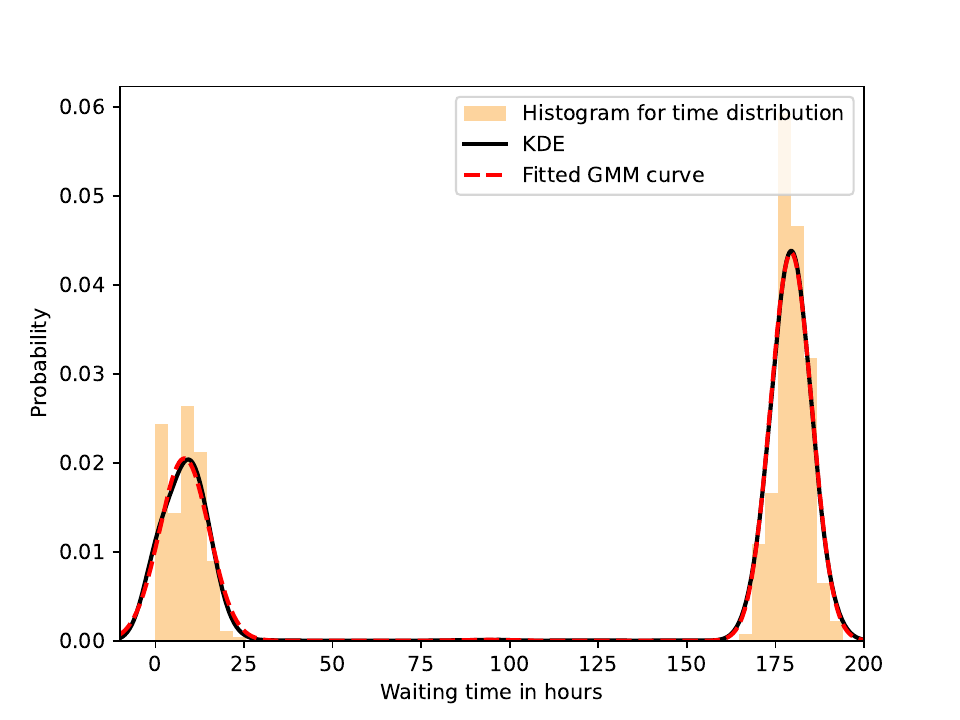}
\caption{PDFs $f_{3,3}(t)$ of the waiting times for the transition $(3,3)$, \ie the self-loop transition of the $\langle\text{{\ttfamily"}Completed{\ttfamily"}}\rangle$ state.}
\label{fig:dist1}
\end{subfigure}%
\centering
\begin{subfigure}{.5\textwidth}
\centering\captionsetup{width=.8\linewidth}
\includegraphics[width=1.0\textwidth]{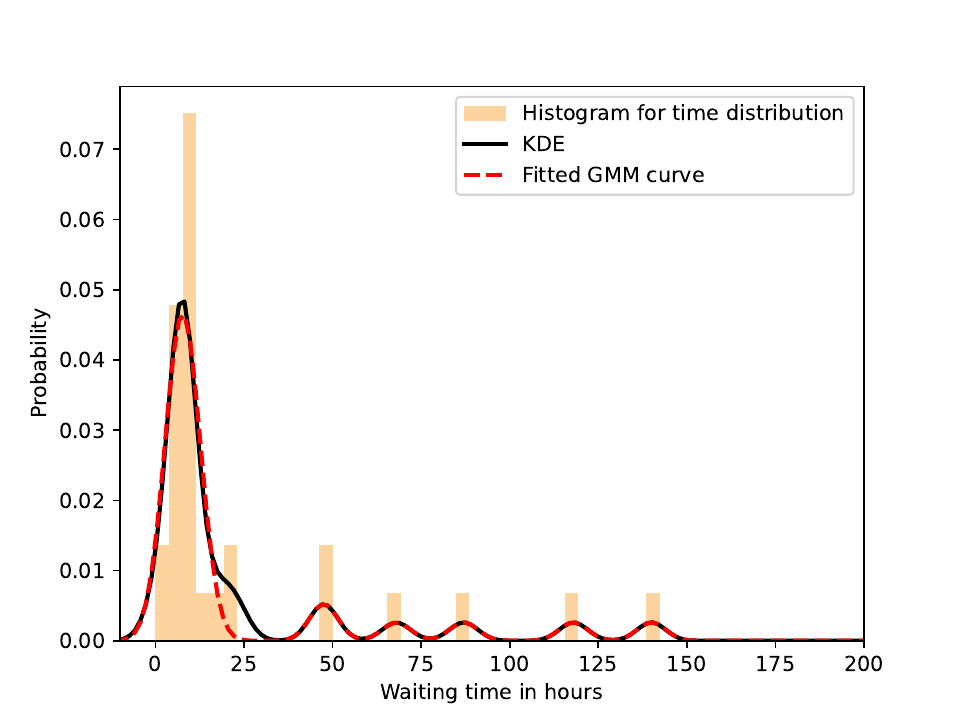}
\caption{PDFs $f_{2,3}(t)$ of the waiting times for the transition between the states $\langle\text{{\ttfamily"}Queued{\ttfamily"}}\rangle$ and $\langle\text{{\ttfamily"}Completed{\ttfamily"}}\rangle$.}
\label{fig:dist2}
\end{subfigure}
\caption{PDFs of the waiting times for the transitions of the semi-Markov process flow presented in \autoref{fig:real_flow}. Discrete PDFs in the form of histograms with 200 bins are presented in light orange. KDE approximations (with Gaussian kernel and bandwidth of~4) and fitted GMMs are black and dashed red, respectively.}
\label{fig:transition_time_dist}
\end{figure*}

Once waiting times for a pair of process states have been collected by \autoref{alg:stochastic:matrix:discovery}, the corresponding time distribution can be derived in a discrete form or in a form of a GMM.

The discrete time distributions can be built from  normalized histograms. For the  event logs analyzed in this paper, we build histograms where each bin corresponds to an hour-wise timestamp.

The GMM is inferred by
approximating the data, using a Kernel Density Estimator (KDE)\footnote{We used \url{https://www.statsmodels.org/stable/} and \url{https://scipy.org/} for KDE and curve fitting.} with a Gaussian kernel to build a smooth and general representation of the PDF.

We then fit a GMM to the KDE representation by optimising the fit of the sum $\sum_{i=1}^\ell q_ig_i(t)$, where $\sum_{i=1}^\ell q_i=1$ and each $g_i(t)$ is a Gaussian with a mean $\mu_i$ and variance $\sigma^2_i$. The result is a set of optimal fitting parameters: $\ell$, $q_i$, $\mu_i$, and $\sigma^2_i$, for $1\leq i\leq \ell$.

Figure~\ref{fig:transition_time_dist} presents histograms of two time distributions and their continuous KDE and GMM approximations: (1)~for the self-loop transition of the state $\langle\text{{\ttfamily"}Completed{\ttfamily"}}\rangle$ (Figure~\ref{fig:dist1}), and, (2) for the transition between the states $\langle\text{{\ttfamily"}Queued{\ttfamily"}}\rangle$ 
and $\langle\text{{\ttfamily"}Completed{\ttfamily"}}\rangle$.
This example demonstrates that durations of events in real-world can be \hl{multi-modal, \emph{i.e.}, contain several local peaks, and} far from exponentially distributed. Hence, regular Markov processes are not always applicable. \hl{More general} models such as semi-Markov processes are needed to capture such distributions with any degree of accuracy.

Note that a GMM permits negative execution times. However, 
it is easy to see that for the GMM that fits to the data shown in \autoref{fig:transition_time_dist} this is an almost trivial issue, as is usually the case.  We will also demonstrate, in the next subsection, that even with the possibility of negative times, accurate performance models can be derived.

Also in \autoref{fig:transition_time_dist} we see how the KDE process smooths the original data. This is a deliberate form of {\em regularization} used to increase the generalizability of results. That is, smoothing helps prevent overfitting to aspects of a particular trace, making conclusions drawn from the process more robust. The GMM model fitting, for instance, is more robust when applied to KDE results, than if applied to raw data.   
However, rather than measuring the quality of this fit, we will show in our results the quality of the inferences obtained from this process. In the next subsection, we will also discuss cases when GMM models can be inferred faster  than discrete models that give precise results but  lack generalizability. 

\subsubsection{Relating Discovered and Observed Process Execution Times}
\label{sec:case_studies}

In this subsection, we apply the proposed full analysis approach to real-world event data including the dataset described above and also including 2 additional datasets. The three logs are from:
\begin{itemize}
    \item the earlier discussed incident management system of Volvo IT Belgium (\emph{BPI'13 incidents}~\cite{WardSteeman2013} event log); and

    \item a university travel expense claim system (\emph{Domestic Declarations--}\allowbreak \emph{DD'20} \cite{https://doi.org/10.4121/uuid:3f422315-ed9d-4882-891f-e180b5b4feb5} and \emph{Request For Payment} --RFP'20~\cite{https://doi.org/10.4121/uuid:895b26fb-6f25-46eb-9e48-0dca26fcd030} event logs).
\end{itemize}

Table~\ref{tab:event_logs_ch} presents the number of traces and events, the maximum support size (the maximum duration in hours one step in the process can take), as well as the number of event names in these event logs. The median trace length for all the event logs is 5.

\begin{table}[h!]
\caption{Characteristics of the real-world event logs.}
\label{tab:event_logs_ch}
\centering 
\begin{tabular}{| c | c | c | c | c |} 
	\hline
\multicolumn{1}{|c}{\scalebox{0.85}{\text{  }Event log\text{  }}}&\multicolumn{1}{|c}{\scalebox{0.85}{\text{  }\# Traces\text{  }}}&\multicolumn{1}{|c}{\scalebox{0.85}{\text{\# Events}\text{  }}}&\multicolumn{1}{|c}{\scalebox{0.85}{\text{Max. support size}\text{  }}}&\multicolumn{1}{|c|}{\scalebox{0.85}{\# Event names}}\\
\hline
\hline

\scalebox{0.85}{BPI\textquotesingle13 inc.}&\scalebox{0.85}{7,554}&\scalebox{0.85}{65,533}&\scalebox{0.85}{17,335}&\scalebox{0.85}{4}\\

\hline

\scalebox{0.85}{DD\textquotesingle20}&\scalebox{0.85}{10,500}&\scalebox{0.85}{56,437}&\scalebox{0.85}{11,263}&\scalebox{0.85}{17}\\

\hline

\scalebox{0.85}{RFP\textquotesingle20}&\scalebox{0.85}{6,886}&\scalebox{0.85}{36,796}&\scalebox{0.85}{9,552}&\scalebox{0.85}{19}\\

\hline
\hline

\end{tabular}

\end{table}

For the analyzed datasets, all activity durations were rounded and estimated in hours, and then corresponding semi-Markov models of order from 1 to 5 were discovered. To estimate the discovery time and the overall GMMs fitting time, we ran 5 experiments for each event log and each order~$k$. Table~\ref{tab:discovery} shows the order, size (number of states and transitions), state average degree,  state maximum degree\footnote{The degrees of the start and end states were not considered because these states were not reduced.}, mean and 95\% CIs (confidence intervals) for the discovery and GMMs fitting times in seconds\footnote{\label{note1}The experiments were run on Intel Xeon w5-3435X × 32, 512 GB.}.
Note that for discrete representations of distributions no fitting was needed and the discrete distributions were derived from histograms directly.
\begin{table}[h!]
%\vspace{-15pt}
%\footnotesize
\caption{Order, size, average state degree, maximum state degree, discovery and overall GMMs fitting times for models constructed from real-world event data.}
\label{tab:discovery}
\centering 
%\normalsize
\begin{tabular}[h!]{| c | c | c | c | c | c |}
	\hline
\multicolumn{1}{|c}{\scalebox{0.85}{\text{  }Event\text{  }}}&\multicolumn{1}{|c}{\scalebox{0.85}{\text{  }Order\text{  }}}&\multicolumn{1}{|c}{\scalebox{0.85}{\text{\# States /}}}&\multicolumn{1}{|c}{\scalebox{0.85}{Avg./Max}}&\multicolumn{1}{|c}{\scalebox{0.85}{\text{  }Disc.time}}&\multicolumn{1}{|c|}{\scalebox{0.85}{Fit.time\text{  }}}\\
\multicolumn{1}{|c}{\scalebox{0.85}{\text{log}\text{  }}}&\multicolumn{1}{|c}{\scalebox{0.85}{\text{  }}}
&\multicolumn{1}{|c}{\scalebox{0.85}{\text{\# Trans.}}}
&\multicolumn{1}{|c}{\scalebox{0.85}{\text{state degree}\text{  }}}
&\multicolumn{1}{|c}{\scalebox{0.85}{\text{  }(in sec.)\text{  }}}&\multicolumn{1}{|c|}{\scalebox{0.85}{\text{  }(in sec.)\text{  }}}\\
\hline
\hline

\scalebox{0.85}{}&\scalebox{0.85}{1}&\scalebox{0.85}{6 / 17}&\scalebox{0.85}{3.00 / 4}&\scalebox{0.85}{0.31$\pm$0.14}&\scalebox{0.85}{11.23$\pm$0.85}\\
\cline{2-6}
\scalebox{0.85}{BPI\textquotesingle13}&\scalebox{0.85}{2}&\scalebox{0.85}{16 / 47}&\scalebox{0.85}{3.14 / 5}&\scalebox{0.85}{0.37$\pm$0.01}&\scalebox{0.85}{13.16$\pm$1.06}\\
\cline{2-6}
\scalebox{0.85}{inc.}&\scalebox{0.85}{3}&\scalebox{0.85}{42 / 105}&\scalebox{0.85}{2.68 / 5}&\scalebox{0.85}{0.33$\pm$0.08}&\scalebox{0.85}{19.40$\pm$0.65}\\
\cline{2-6}
\scalebox{0.85}{ }&\scalebox{0.85}{4}&\scalebox{0.85}{91 / 219}&\scalebox{0.85}{2.65 / 5}&\scalebox{0.85}{0.36$\pm$0.06}&\scalebox{0.85}{33.15$\pm$0.11}\\
\cline{2-6}
\scalebox{0.85}{ }&\scalebox{0.85}{5}&\scalebox{0.85}{191 / 432}&\scalebox{0.85}{2.56 / 5}&\scalebox{0.85}{0.39$\pm$0.01}&\scalebox{0.85}{88.08$\pm$1.41}\\
\hline
\hline

\scalebox{0.85}{}&\scalebox{0.85}{1}&\scalebox{0.85}{19 / 48}&\scalebox{0.85}{3.12 / 10}&\scalebox{0.85}{0.29$\pm$0.05}&\scalebox{0.85}{34.98$\pm$5.52}\\
\cline{2-6}
\scalebox{0.85}{}&\scalebox{0.85}{2}&\scalebox{0.85}{43 / 90}&\scalebox{0.85}{2.46 / 8}&\scalebox{0.85}{0.29$\pm$0.08}&\scalebox{0.85}{183.84$\pm$15.05}\\
\cline{2-6}
\scalebox{0.85}{DD\textquotesingle20}&\scalebox{0.85}{3}&\scalebox{0.85}{75 /139}&\scalebox{0.85}{2.14 / 8}&\scalebox{0.85}{0.30$\pm$0.17}&\scalebox{0.85}{185.97$\pm$17.80}\\
\cline{2-6}
\scalebox{0.85}{ }&\scalebox{0.85}{4}&\scalebox{0.85}{118 / 197}&\scalebox{0.85}{1.90 / 8}&\scalebox{0.85}{0.34$\pm$0.18}&\scalebox{0.85}{421.31$\pm$20.30}\\
\cline{2-6}
\scalebox{0.85}{ }&\scalebox{0.85}{5}&\scalebox{0.85}{168 / 256}&\scalebox{0.85}{1.72 / 8}&\scalebox{0.85}{0.40$\pm$0.10}&\scalebox{0.85}{424.50$\pm$18.00}\\
\hline
\hline

\scalebox{0.85}{}&\scalebox{0.85}{1}&\scalebox{0.85}{21 /57}&\scalebox{0.85}{3.26 / 9}&\scalebox{0.85}{0.22$\pm$0.16}&\scalebox{0.85}{23.17$\bf\pm$1.49}\\
\cline{2-6}
\scalebox{0.85}{}&\scalebox{0.85}{2}&\scalebox{0.85}{48 / 99}&\scalebox{0.85}{2.33 /7}&\scalebox{0.85}{0.21$\pm$0.06}&\scalebox{0.85}{25.33$\pm$0.89}\\
\cline{2-6}
\scalebox{0.85}{RFP\textquotesingle20}&\scalebox{0.85}{3}&\scalebox{0.85}{81 / 151}&\scalebox{0.85}{2.10 / 7}&\scalebox{0.85}{0.21$\pm$0.06}&\scalebox{0.85}{25.34$\pm$0.89}\\
\cline{2-6}
\scalebox{0.85}{ }&\scalebox{0.85}{4}&\scalebox{0.85}{124 / 204}&\scalebox{0.85}{1.86 / 7}&\scalebox{0.85}{0.20$\pm$0.06}&\scalebox{0.85}{30.58$\pm$2.85}\\
\cline{2-6}
\scalebox{0.85}{ }&\scalebox{0.85}{5}&\scalebox{0.85}{169 / 255}&\scalebox{0.85}{1.69 / 7}&\scalebox{0.85}{0.24$\pm$0.05}&\scalebox{0.85}{31.38$\pm$0.81}\\
\hline
\hline

\end{tabular}

\end{table}

As the results suggest, the discovery algorithm is computationally negligible, but it takes more time to find all GMMs to annotate transitions of the discovered models. As studied in \cite{pmlr-v35-daskalakis14,DBLP:journals/corr/LiS15a}, the fitting complexity is exponential on the number of Gaussians in a GMM. Moreover, the larger the discovered model is (the more transitions it contains), the more time is needed to build all GMMs. However, in the case of \emph{RFP'20} event log, durations of transitions of higher-order models can be approximated faster, this can be explained by the fact that the optimizing fitting algorithm can converge differently depending on the input data and the number of initially guessed fitted parameters.

\begin{figure}
\vspace{-20pt}
\centering\captionsetup{width=1.0\linewidth}
\includegraphics[width=0.5\textwidth]{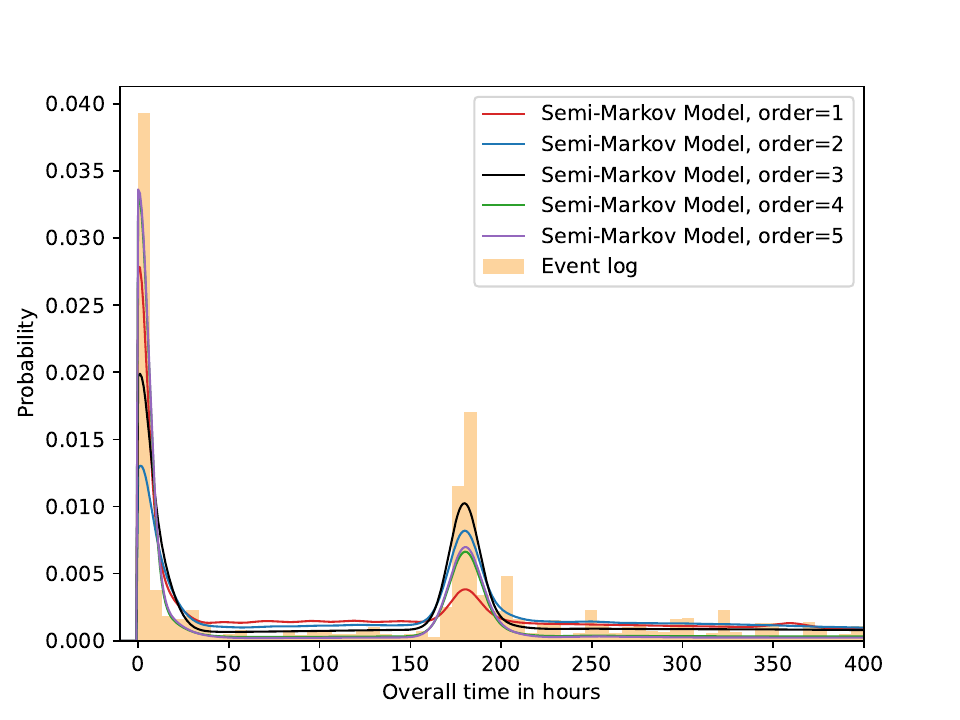}
\caption{The overall process execution time for the incident management system event log (BPI'13 incidents) and semi-Markov models with continuous GMM distributions discovered from this event
log with weight threshold=0.001.}
\label{fig:dist_bpi}
\end{figure}

To relate the observed and discovered process execution times we plotted the time distributions for the \emph{BPI'13 incidents} event log and semi-Markov models discovered from this event log.
Figure~\ref{fig:dist_bpi} shows the overall time distributions for the event log compared to the estimated times for semi-Markov models of order $k$ from 1 to 5 with continuous GMM distributions discovered from this event log setting the \textbf{weight threshold} to 0.001 (all the Gaussians with weight coefficients less than this threshold were merged into one Gaussian).

Gaussian distributions of the final models were replaced by the corresponding truncated Gaussian distributions to avoid negative times. As can be observed from the plots, the discovered models allow location of the main peaks of time distributions. However, they represent a more general and smooth behavior.

To calculate the \emph{(performance) accuracy} of the discovered continious (GMM) models we used Kullback–Leibler divergence.  Kullback–Leibler divergence (KL divergence)~\cite{Kullback1951OnIA} is a standard measure used in machine learning algorithms for evaluating  the models. For each discovered process model and a corresponding event log we calculated the KL divergence~as:
$$ KL(L,M)=\sum\limits_{t\in T} L(t)\ \ln\Big(\frac{L(t)}{M(t)}\Big)\ ,$$
where $t$ iterates over the set of bins (time intervals) $T$, $L(t)$ is the observed probability (obtained directly from the event log) and $M(t)$ is the discovered probability of being in time interval $t$. This divergence relates time distributions and represents the expected logarithmic difference between the observed and modeled probabilities.

\autoref{tab:gmm_1} and \autoref{tab:gmm_2} show the KL divergence, \hl{average size of GMMs, \emph{i.e.}, the average number of components (Gaussians) in the mixture models}, and the mean  and 95\% CI for the reduction time of GMM semi-Markov models built from the real-world event data with the weight thresholds set to 0.001 and 0.0001, respectively.
KL divergence was computed for 20 equal intervals spanning all process times up to 1,200 hours.  For each event log, each order $k$, and each threshold we ran 5 experiments. When the models were not discovered, the corresponding values were not provided.

\begin{table}[h!]
%\vspace{-15pt}
\caption{\hl{KL divergence (accuracy), average size of GMMs and reduction time for the models constructed from real-world event data. Continuous (GMM) case, weight threshold=0.001.}}
\label{tab:gmm_1}
\centering 
\begin{tabular}[h!]{| c | c | c | c | c | c |}
	\hline
\multicolumn{1}{|c}{\scalebox{0.85}{\text{  }Event log\text{  }}}&\multicolumn{1}{|c}{\scalebox{0.85}{\text{  }Order $k$\text{  }}}&\multicolumn{1}{|c}{\scalebox{0.85}{\text{\text{  }KL divergence\text{  }}\text{  }}}&\multicolumn{1}{|c}{\scalebox{0.85}{\text{\text{  }\hl{Avg. size of}\text{  }}\text{  }}}&\multicolumn{1}{|c|}{\scalebox{0.85}{Reduct. time}}\\
\multicolumn{1}{|c }{\scalebox{0.85}{\text{  }\text{  }}}&\multicolumn{1}{|c}{\scalebox{0.85}{\text{  }\text{  }}}&\multicolumn{1}{|c}{\scalebox{0.85}{\text{\text{  }\text{  }}\text{  }}}&\multicolumn{1}{|c}{\scalebox{0.85}{\text{\text{  } \hl{GMMs}\text{  }}\text{  }}}&\multicolumn{1}{|c|}{\scalebox{0.85}{(in sec.)}}\\
\hline
\hline

\scalebox{0.85}{}&\scalebox{0.85}{1}&\scalebox{0.85}{0.0930}&\scalebox{0.85}{42.33}&\scalebox{0.85}{0.02$\pm$0.00}\\
\cline{2-5}
\scalebox{0.85}{BPI\textquotesingle13}&\scalebox{0.85}{2}&\scalebox{0.85}{0.0280}&\scalebox{0.85}{32.17}&\scalebox{0.85}{0.45$\pm$0.03}\\
\cline{2-5}
\scalebox{0.85}{incidents}&\scalebox{0.85}{\textbf{3}}&\scalebox{0.85}{\textbf{0.0228}}&\scalebox{0.85}{\textbf{54.83}}&\scalebox{0.85}{\textbf{2.29}$\bf\pm$\textbf{0.07}}\\
\cline{2-5}
\scalebox{0.85}{ }&\scalebox{0.85}{4}&\scalebox{0.85}{0.1554}&\scalebox{0.85}{76.34}&\scalebox{0.85}{14.27$\pm$0.76}\\
\cline{2-5}
\scalebox{0.85}{ }&\scalebox{0.85}{5}&\scalebox{0.85}{0.5105}&\scalebox{0.85}{89.73}&\scalebox{0.85}{70.30$\pm$1.43}\\
\hline
\hline

\scalebox{0.85}{}&\scalebox{0.85}{\textbf{1}}&\scalebox{0.85}{\textbf{0.0930}}&\scalebox{0.85}{\textbf{76.56}}&\scalebox{0.85}{\textbf{0.11}$\pm$\textbf{0.13}}\\
\cline{2-5}
\scalebox{0.85}{}&\scalebox{0.85}{2}&\scalebox{0.85}{0.1458}&\scalebox{0.85}{92.93}&\scalebox{0.85}{0.96$\pm$0.27}\\
\cline{2-5}
\scalebox{0.85}{DD\textquotesingle20}&\scalebox{0.85}{3}&\scalebox{0.85}{0.1299}&\scalebox{0.85}{79.09}&\scalebox{0.85}{1.93$\bf\pm$0.13}\\
\cline{2-5}
\scalebox{0.85}{ }&\scalebox{0.85}{4}&\scalebox{0.85}{0.1568}&\scalebox{0.85}{77.83}&\scalebox{0.85}{2.11$\pm$0.79}\\
\cline{2-5}
\scalebox{0.85}{ }&\scalebox{0.85}{5}&\scalebox{0.85}{0.1448}&\scalebox{0.85}{71.95}&\scalebox{0.85}{1.62$\pm$0.19}\\
\hline
\hline

\scalebox{0.85}{}&\scalebox{0.85}{\textbf{1}}&\scalebox{0.85}{\textbf{0.0779}}&\scalebox{0.85}{\textbf{80.66}}&\scalebox{0.85}{\textbf{0.06}$\pm$\textbf{0.00}}\\
\cline{2-5}
\scalebox{0.85}{}&\scalebox{0.85}{2}&\scalebox{0.85}{0.0918}&\scalebox{0.85}{81.17}&\scalebox{0.85}{0.46
$\pm$0.05}\\
\cline{2-5}
\scalebox{0.85}{RFP\textquotesingle20}&\scalebox{0.85}{3}&\scalebox{0.85}{0.1209}&\scalebox{0.85}{81.15}&\scalebox{0.85}{1.26$\pm$0.06}\\
\cline{2-5}
\scalebox{0.85}{ }&\scalebox{0.85}{4}&\scalebox{0.85}{0.1162}&\scalebox{0.85}{69.68}&\scalebox{0.85}{1.34$\pm$0.18}\\
\cline{2-5}
\scalebox{0.85}{ }&\scalebox{0.85}{5}&\scalebox{0.85}{0.1490}&\scalebox{0.85}{74.67}&\scalebox{0.85}{1.67$\pm$0.10}\\
\hline
\hline

\end{tabular}

\end{table}

\begin{table}[h!]

\caption{\hl{KL divergence (accuracy), average size of GMMs and reduction time for the models constructed from real-world event data. Continuous (GMM) case, weight threshold=0.0001.}}
\label{tab:gmm_2}
\centering 
\begin{tabular}[h!]{| c | c | c | c | c |}
	\hline
\multicolumn{1}{|c}{\scalebox{0.85}{\text{  }Event log\text{  }}}&\multicolumn{1}{|c}{\scalebox{0.85}{\text{  }Order $k$\text{  }}}&\multicolumn{1}{|c}{\scalebox{0.85}{\text{\text{  }KL divergence\text{  }}\text{  }}}&\multicolumn{1}{|c}{\scalebox{0.85}{\text{\text{  }\hl{Avg. size of}\text{  }}\text{  }}}&\multicolumn{1}{|c|}{\scalebox{0.85}{Reduct. time}}\\
\multicolumn{1}{|c }{\scalebox{0.85}{\text{  }\text{  }}}&\multicolumn{1}{|c}{\scalebox{0.85}{\text{  }\text{  }}}&\multicolumn{1}{|c}{\scalebox{0.85}{\text{\text{  }\text{  }}\text{  }}}&\multicolumn{1}{|c}{\scalebox{0.85}{\text{\text{  } \hl{GMMs}\text{  }}\text{  }}}&\multicolumn{1}{|c|}{\scalebox{0.85}{(in sec.)}}\\
\hline
\hline

\scalebox{0.85}{}&\scalebox{0.85}{1}&\scalebox{0.85}{0.0882}&\scalebox{0.85}{173.58}&\scalebox{0.85}{0.55$\pm$0.08}\\
\cline{2-5}
\scalebox{0.85}{BPI\textquotesingle13}&\scalebox{0.85}{\textbf{2}}&\scalebox{0.85}{\textbf{0.0038}}&\scalebox{0.85}{\textbf{250.84}}&\scalebox{0.85}{\textbf{580.30}$\bf\pm$\textbf{6.84}}\\
\cline{2-5}
\scalebox{0.85}{incidents}&\scalebox{0.85}{3}&\scalebox{0.85}{0.0104}&\scalebox{0.85}{191.64}&\scalebox{0.85}{1,104.25$\pm$17.60}\\
\cline{2-5}
\scalebox{0.85}{ }&\scalebox{0.85}{4}&\scalebox{0.85}{--}&\scalebox{0.85}{--}&\scalebox{0.85}{--}\\
\cline{2-5}
\scalebox{0.85}{ }&\scalebox{0.85}{5}&\scalebox{0.85}{--}&\scalebox{0.85}{--}&\scalebox{0.85}{--}\\
\hline
\hline

\scalebox{0.85}{}&\scalebox{0.85}{\textbf{1}}&\scalebox{0.85}{\textbf{0.0543}}&\scalebox{0.85}{\textbf{365.05}}&\scalebox{0.85}{\textbf{3.18}$\pm$\textbf{0.08}}\\
\cline{2-5}
\scalebox{0.85}{}&\scalebox{0.85}{2}&\scalebox{0.85}{0.1048}&\scalebox{0.85}{340.16}&\scalebox{0.85}{659.45$\pm$23.72}\\
\cline{2-5}
\scalebox{0.85}{DD\textquotesingle20}&\scalebox{0.85}{3}&\scalebox{0.85}{0.1127}&\scalebox{0.85}{346.31}&\scalebox{0.85}{1,425.97$\bf\pm$57.06}\\
\cline{2-5}
\scalebox{0.85}{ }&\scalebox{0.85}{4}&\scalebox{0.85}{0.1079}&\scalebox{0.85}{330.38}&\scalebox{0.85}{1,925.59$\pm$40.72}\\
\cline{2-5}
\scalebox{0.85}{ }&\scalebox{0.85}{5}&\scalebox{0.85}{0.1137}&\scalebox{0.85}{352.80}&\scalebox{0.85}{1,630.18$\pm$13.50}\\
\hline
\hline

\scalebox{0.85}{}&\scalebox{0.85}{\textbf{1}}&\scalebox{0.85}{\textbf{0.0753}}&\scalebox{0.85}{\textbf{355.00}}&\scalebox{0.85}{\textbf{0.72}$\pm$\textbf{0.05}}\\
\cline{2-5}
\scalebox{0.85}{}&\scalebox{0.85}{2}&\scalebox{0.85}{0.0915}&\scalebox{0.85}{313.65}&\scalebox{0.85}{397.06$\pm$15.53}\\
\cline{2-5}
\scalebox{0.85}{RFP\textquotesingle20}&\scalebox{0.85}{3}&\scalebox{0.85}{0.0950}&\scalebox{0.85}{323.10}&\scalebox{0.85}{641.61$\pm$33.93}\\
\cline{2-5}
\scalebox{0.85}{ }&\scalebox{0.85}{4}&\scalebox{0.85}{0.0860}&\scalebox{0.85}{326.34}&\scalebox{0.85}{868.09$\pm$36.29}\\
\cline{2-5}
\scalebox{0.85}{ }&\scalebox{0.85}{5}&\scalebox{0.85}{0.0975}&\scalebox{0.85}{322.55}&\scalebox{0.85}{973.40$\pm$33.64}\\
\hline
\hline

\end{tabular}

\end{table}

Because of the pruning technique, the order in which states are reduced can affect the final distribution. To mitigate this, we applied a heuristic and first removed states with the minimal $m_1\cdot m_2$ value, where $m_1$ is the number incoming transitions and $m_2$ is the number of outgoing transitions of a state. Additionally, removing less connected states first, can reduce the number of convolution operations and as a result get better accuracy.

In \autoref{tab:gmm_1} and \autoref{tab:gmm_2}, the best models in terms of accuracy for each of the event logs and each of the thresholds are highlighted in bold. Although, one might expect that the accuracy should increase as the order of discovered models increases, these results show that this is not always the case. This can be explained by the fact that \hl{durations are rounded to hours before applying convolutions, so in large higher-order models, this initial rounding of transition durations can reduce accuracy.}

\begin{figure}[t!]
%\vspace{-20pt}
\centering\captionsetup{width=1.\linewidth}
\includegraphics[width=0.5\textwidth]{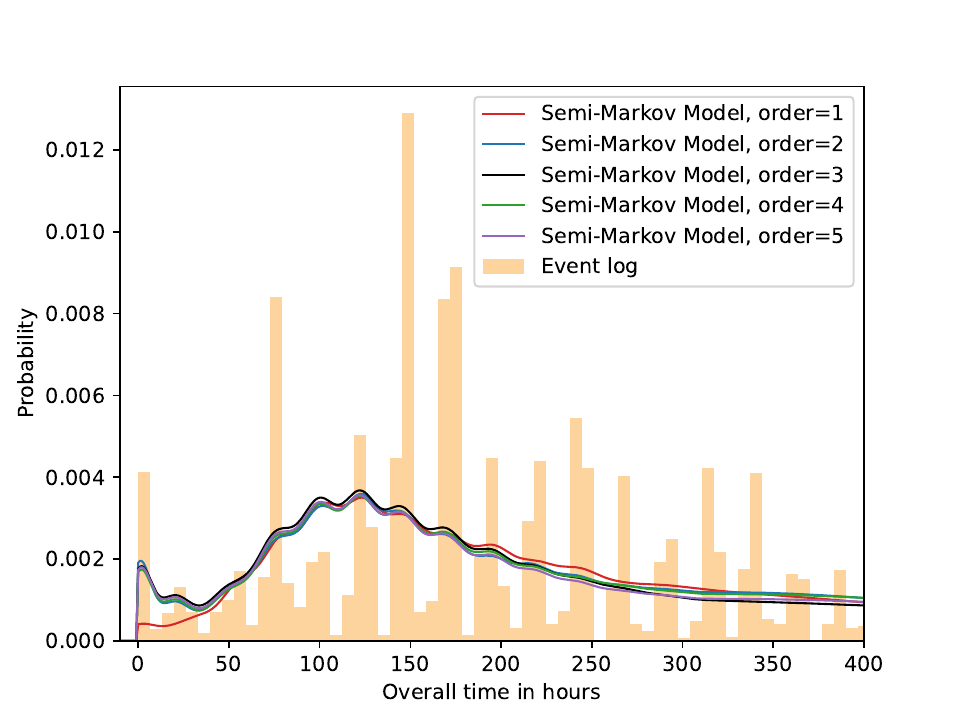}
\caption{The overall process execution time for the university
travel expense claim system event log (RFP’20) and semi-Markov models with continuous time GMM distributions discovered from this event log with weight threshold=0.001.}
\label{fig:dist_3}
\end{figure}

\begin{figure}[t!]
\centering\captionsetup{width=1.\linewidth}
%\vspace{-15pt}
\includegraphics[width=0.5\textwidth]{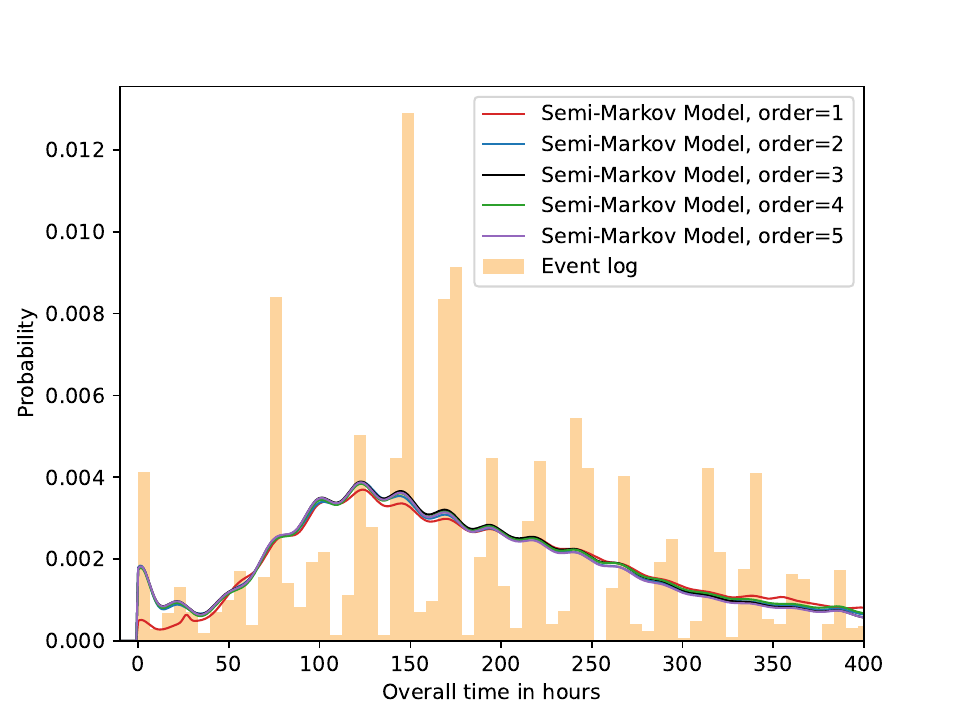}
\caption{The overall execution time for RFP’20 event log and semi-Markov models with continuous GMM time distributions discovered from this event log with weight threshold=0.0001.}
\label{fig:dist_4}
\end{figure}

\autoref{fig:dist_3}  and \autoref{fig:dist_4} present the results for models discovered from \emph{RFP'20} event log with different thresholds.  The models discovered with the weight threshold set to 0.0001 (Figure~\ref{fig:dist_4}) capture more peaks than the models discovered with the 0.001 weight threshold (Figure~\ref{fig:dist_3}), because they are more accurate and can better represent longer traces.

\begin{figure}[t!]
%\vspace{-21pt}
%\begin{center}
\centering\includegraphics[width=0.5\textwidth]{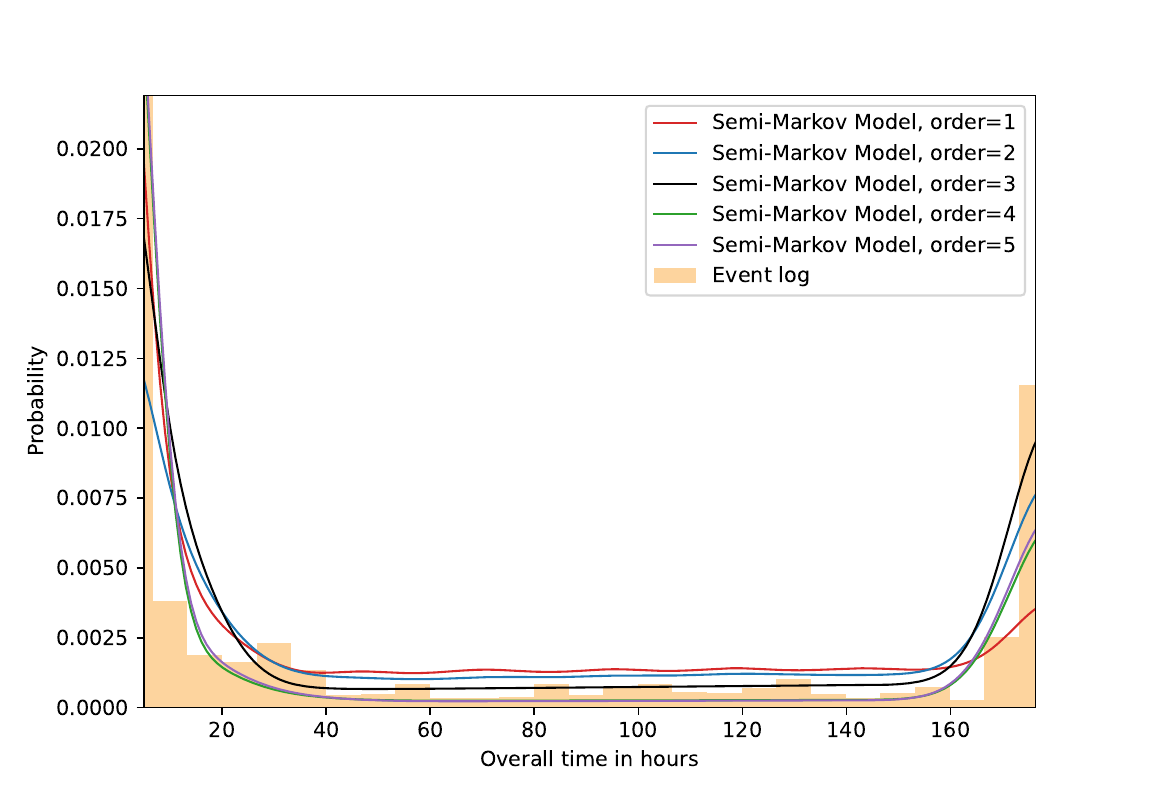}
\centering\caption{Zoomed-in time distributions for  semi-Markov models with continuous time distributions discovered from \emph{BPI'13 incidents} event log with the threshold set to 0.001 (see the original Figure~\ref{fig:dist_bpi}).}
\label{fig:zoom_in}
%\end{center}
\vspace{-11pt}
\end{figure}

One more observation is that for \emph{BPI'13 incidents} event log the accuracy drops for $k$ larger than 3.
In this case, the total number of Gaussians associated with all the transitions of higher-order models is large, and hence the pruning technique that reduces the accuracy will be  applied at the last steps of the reduction. 
This results in higher-order models approximating the main peaks but not the minor time variations. As can be observed from Figure~\ref{fig:zoom_in} (Figure~\ref{fig:zoom_in} zooms in on the time distributions presented in Figure~\ref{fig:dist_bpi}), the distributions for semi-Markov models of order 4 and 5 are flatter than for other models and capture only the main peaks.

\hl{Reducing large higher-order models takes more time. Additionally, as indicated by the complexity analysis (\mbox{\autoref{sec:gaussians}}) and the experimental results (\mbox{\autoref{tab:gmm_1} and \autoref{tab:gmm_2}}), the lower the threshold -- or, in other words, the larger the size of the GMMs -- the slower the computations become. The complexity analysis (\mbox{\autoref{sec:gaussians}}) also shows that the time complexity is quadratic with respect to the number of GMM components. Therefore, the current implementation could be further optimized by selecting more efficient data structures. We consider this a potential direction for future work.}

The combination of accuracy and computation times suggest that small values of $k$ between 1 and 2 are good choices for 0.001 and 0.0001 thresholds. The best choice for a particular application will vary. But in most cases these smaller values were, if not optimal, not far from. 

These experiments demonstrate the applicability of the proposed full continuous GMM-based analysis technique. It is shown that the best accuracy is in the range from 0.0038  to 0.0753 (the expected relations of observed and modeled processes' durations are in the interval from 1.0038 to 1.0782), and the most accurate overall time PDFs can be discovered in a reasonable amount of time (from less than 1 second to 9.7~minutes).

\begin{figure}[h!]
\centering\captionsetup{width=1.0\linewidth}
\includegraphics[width=0.5\textwidth]{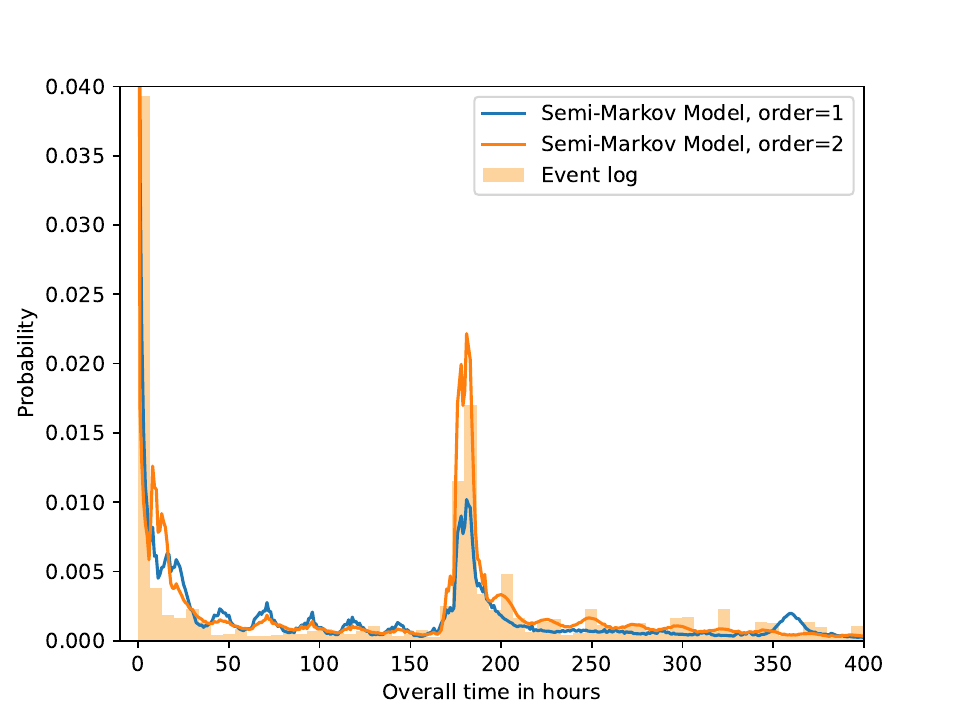}
\caption{The overall process execution time for the incident management system event log (BPI'13 incidents) and semi-Markov models with discrete time distributions discovered from this event
log.}
\label{fig:discrete_bpi}
\end{figure}

\begin{figure}[h!]
\centering\captionsetup{width=1.0\linewidth}
%\vspace{-15pt}
\includegraphics[width=0.5\textwidth]{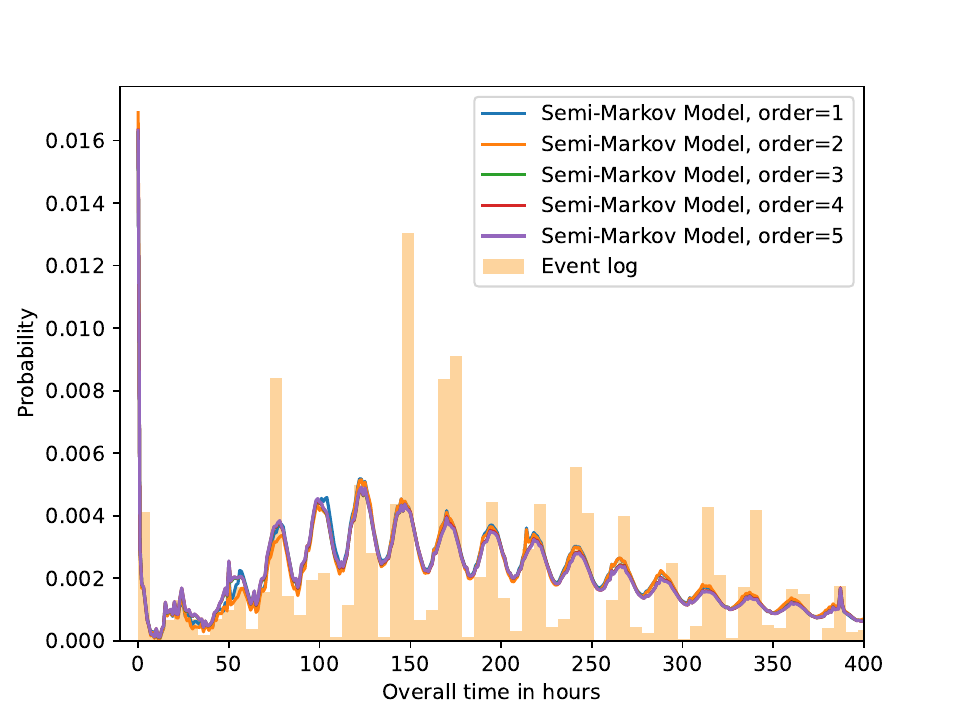}
\caption{The overall process execution time for RFP’20 event log and discovered semi-Markov models with discrete time distributions.}
\label{fig:dist_5}
\end{figure}

Another approach implemented in this work is based on discrete distributions built from the histograms associated with the models' transitions. 
The overall time distributions for semi-Markov models \hl{discovered from \mbox{\emph{BPI'13 incidents}} and \mbox{\emph{RFP'20}} event logs using the discrete approach are presented in~\mbox{\autoref{fig:discrete_bpi}} and \mbox{\autoref{fig:dist_5}}, respectively}. Due to the computational complexity and the size of support, only the first and second order discrete models could be discovered \hl{from \mbox{\emph{BPI'13 incidents}}}. These models are more precise and less general and are represented by finite discrete distributions with large support. 
Note that for both approaches the \textbf{loop threshold} was set to 0.1. This value was selected empirically because it provided the same outcomes as the more precise approaches (with a smaller \textbf{loop threshold}), but in a shorter~time.

\autoref{tab:discrete} presents average support size for the transitions of models (constructed \hl{during the reduction}) and the reduction time for these models. Although the overall results for the discrete approach are promising \hl{(see the KL divergence for the discrete approach in~\mbox{\autoref{tab:kl_discrete_sim}})}, for  \emph{BPI'13 incidents} event log, reduction, even when executed using optimized code for calculating convolutions from NumPy library\footnote{https://numpy.org/}, cannot be performed for larger orders due to the size of the supports and the number of convolution operations related to the number of states (see \autoref{tab:event_logs_ch}, \autoref{tab:discovery}\hl{, and \mbox{\autoref{tab:discrete}}}). Note that the size of supports increases with each convolution and hence with each node reduction.
In such cases, the continuous GMM approach with a weight threshold set to 0.001 works faster. Another drawback of the discrete technique is that it is not suitable for what-if analysis, where time distributions are usually represented analytically rather than as large arrays of numbers~\hl{\mbox{\autoref{tab:discrete}}}.
\begin{table}[h!]

\caption{Average~size of support and reduction time for models constructed from real-world event data.  Discrete case.}
\label{tab:discrete}
\centering 
\begin{tabular}[h!]{| c | c | c | c |}
	\hline
\multicolumn{1}{|c}{\scalebox{0.85}{\text{  }Event log\text{  }}}&\multicolumn{1}{|c}{\scalebox{0.85}{Order $k$}}&\multicolumn{1}{|c}{\scalebox{0.85}{\hl{Avg.~size of}}}&\multicolumn{1}{|c|}{\scalebox{0.85}{Reduct. time}}\\
\multicolumn{1}{|c}{\scalebox{0.85}{\text{  }\text{  }}}&\multicolumn{1}{|c}{\scalebox{0.85}{}}&\multicolumn{1}{|c}{\scalebox{0.85}{\hl{support}}}&\multicolumn{1}{|c|}{\scalebox{0.85}{(in sec.)}}\\
\hline
\hline

\scalebox{0.85}{}&\scalebox{0.85}{1}&\scalebox{0.85}{46,367.58}&\scalebox{0.85}{14.03$\pm$0.30}\\
\cline{2-4}
\scalebox{0.85}{BPI\textquotesingle13}&\scalebox{0.85}{2}&\scalebox{0.85}{27,675.67}&\scalebox{0.85}{252.43$\bf\pm$0.75}\\
\cline{2-4}
\scalebox{0.85}{incidents}&\scalebox{0.85}{3}&\scalebox{0.85}{--}&\scalebox{0.85}{--}\\
\cline{2-4}
\scalebox{0.85}{ }&\scalebox{0.85}{4}&\scalebox{0.85}{--}&\scalebox{0.85}{--}\\
\cline{2-4}
\scalebox{0.85}{ }&\scalebox{0.85}{5}&\scalebox{0.85}{--}&\scalebox{0.85}{--}\\
\hline
\hline

\scalebox{0.85}{}&\scalebox{0.85}{1}&\scalebox{0.85}{8,793.11}&\scalebox{0.85}{0.41$\pm$0.01}\\
\cline{2-4}
\scalebox{0.85}{}&\scalebox{0.85}{2}&\scalebox{0.85}{6,978.79}&\scalebox{0.85}{0.82$\pm$0.19}\\
\cline{2-4}
\scalebox{0.85}{DD\textquotesingle20}&\scalebox{0.85}{3}&\scalebox{0.85}{8,153.62}&\scalebox{0.85}{12.85$\bf\pm$0.19}\\
\cline{2-4}
\scalebox{0.85}{ }&\scalebox{0.85}{4}&\scalebox{0.85}{5,057.89}&\scalebox{0.85}{1.46$\pm$0.88}\\
\cline{2-4}
\scalebox{0.85}{ }&\scalebox{0.85}{5}&\scalebox{0.85}{4,202.99}&\scalebox{0.85}{1.05$\pm$0.03}\\
\hline
\hline

\scalebox{0.85}{}&\scalebox{0.85}{1}&\scalebox{0.85}{8,295.63}&\scalebox{0.85}{0.42$\pm$0.02}\\
\cline{2-4}
\scalebox{0.85}{}&\scalebox{0.85}{2}&\scalebox{0.85}{6,279.02}&\scalebox{0.85}{1.39$\pm$0.16}\\
\cline{2-4}
\scalebox{0.85}{RFP\textquotesingle20}&\scalebox{0.85}{3}&\scalebox{0.85}{7,333.21}&\scalebox{0.85}{2.25$\pm$0.57}\\
\cline{2-4}
\scalebox{0.85}{ }&\scalebox{0.85}{4}&\scalebox{0.85}{4,305.32}&\scalebox{0.85}{1.20$\pm$0.25}\\
\cline{2-4}
\scalebox{0.85}{ }&\scalebox{0.85}{5}&\scalebox{0.85}{4,214.60}&\scalebox{0.85}{1.02$\pm$0.02}\\
\hline
\hline

\end{tabular}

\end{table}

Another observation is that according to the plots (see \autoref{fig:dist_3}, \autoref{fig:dist_4}, \autoref{fig:dist_5}) both approaches tend to be more precise for smaller process durations. This can be explained by the fact that smaller durations correspond to less number of convolutions and, hence, to less number of approximations in both cases.

\hl{To justify the continuous (GMM-based) and discrete approaches, we implemented a semi-Markov model simulation technique. A 10-second simulation was selected, as it yielded results comparable to those obtained from longer-running simulations on the same computational architecture\mbox{\footnoteref{note1}}. Notably, the initial rounding of time durations to hours prevents the KL divergence parameter from converging to zero, even as the simulation time increases. As presented in \mbox{\autoref{tab:kl_discrete_sim}}, the KL divergence values for the discrete approach (where it can be calculated) and the simulation method are closely aligned, representing the optimal outcomes given the rounding constraints. The discrete approach exhibits a computational advantage for smaller support sizes, offering faster performance than the simulation. The continuous GMM approach surpasses both discrete and simulation techniques in terms of representational efficiency. As follows from \mbox{\autoref{tab:gmm_1}, \autoref{tab:gmm_2}, and \autoref{tab:discrete}}, the number of GMMs representing durations is considerably smaller than the support size. Hence, the continuous GMM-based approach enhances model compactness and interpretability, facilitating applications such as \emph{what-if} analysis, duration-based process clustering, and classification.}

In our future work, we plan to develop a hybrid approach that combines continuous and discrete techniques. This approach will aim to improve accuracy and time complexity by applying discretization and curve fitting at each node reduction. This will enable us to limit the number of model parameters while preserving its accuracy. Another direction for the future work is the application of divide-conquer techniques and advanced data structures to improve the precision and \hl{reduce calculation times} for long traces and large number of convolutions.

\begin{table}[h!]
\centering 
\parbox{\linewidth}{
\centering 
\caption{KL divergences of the overall process execution times: the discrete approach and the 10-second simulation technique.}
\label{tab:kl_discrete_sim}

\begin{tabular}[h!]{| c | c | c | c |}
	\hline
\multicolumn{1}{|c}{\scalebox{0.85}{\text{  }Event log\text{  }}}&\multicolumn{1}{|c}{\scalebox{0.85}{Order $k$}}&\multicolumn{1}{|c}{\scalebox{0.85}{KL divergence.}}&\multicolumn{1}{|c|}{\scalebox{0.85}{KL divergence.}}\\
\multicolumn{1}{|c}{\scalebox{0.85}{\text{  }\text{  }}}&\multicolumn{1}{|c}{\scalebox{0.85}{}}&\multicolumn{1}{|c}{\scalebox{0.85}{Discrete case}}&\multicolumn{1}{|c|}{\scalebox{0.85}{Simulation}}\\
\hline
\hline

\scalebox{0.85}{}&\scalebox{0.85}{1}&\scalebox{0.85}{0.1442}&\scalebox{0.85}{0.0834}\\
\cline{2-4}
\scalebox{0.85}{BPI\textquotesingle13}&\scalebox{0.85}{2}&\scalebox{0.85}{0.0510}&\scalebox{0.85}{0.0467}\\
\cline{2-4}
\scalebox{0.85}{incidents}&\scalebox{0.85}{3}&\scalebox{0.85}{--}&\scalebox{0.85}{0.0700}\\
\cline{2-4}
\scalebox{0.85}{ }&\scalebox{0.85}{4}&\scalebox{0.85}{--}&\scalebox{0.85}{0.1234}\\
\cline{2-4}
\scalebox{0.85}{ }&\scalebox{0.85}{5}&\scalebox{0.85}{--}&\scalebox{0.85}{0.1267}\\
\hline
\hline

\scalebox{0.85}{}&\scalebox{0.85}{1}&\scalebox{0.85}{0.0114}&\scalebox{0.85}{0.0077}\\
\cline{2-4}
\scalebox{0.85}{}&\scalebox{0.85}{2}&\scalebox{0.85}{0.0075}&\scalebox{0.85}{0.0082}\\
\cline{2-4}
\scalebox{0.85}{DD\textquotesingle20}&\scalebox{0.85}{3}&\scalebox{0.85}{0.0076}&\scalebox{0.85}{0.0081}\\
\cline{2-4}
\scalebox{0.85}{ }&\scalebox{0.85}{4}&\scalebox{0.85}{0.0075}&\scalebox{0.85}{0.0077}\\
\cline{2-4}
\scalebox{0.85}{ }&\scalebox{0.85}{5}&\scalebox{0.85}{0.0077}&\scalebox{0.85}{0.0082}\\
\hline
\hline

\scalebox{0.85}{}&\scalebox{0.85}{1}&\scalebox{0.85}{0.0129}&\scalebox{0.85}{0.0119}\\
\cline{2-4}
\scalebox{0.85}{}&\scalebox{0.85}{2}&\scalebox{0.85}{0.0088}&\scalebox{0.85}{0.0105}\\
\cline{2-4}
\scalebox{0.85}{RFP\textquotesingle20}&\scalebox{0.85}{3}&\scalebox{0.85}{0.0082}&\scalebox{0.85}{0.0095}\\
\cline{2-4}
\scalebox{0.85}{ }&\scalebox{0.85}{4}&\scalebox{0.85}{0.0090}&\scalebox{0.85}{0.0110}\\
\cline{2-4}
\scalebox{0.85}{ }&\scalebox{0.85}{5}&\scalebox{0.85}{0.0092}&\scalebox{0.85}{0.0109}\\
\hline
\hline

\end{tabular}

}

\end{table}

\section{Conclusion}
\label{sec:conclusion}
The paper proposes express and full techniques for the performance analysis of stochastic process models discovered from event logs. These new techniques comparing to the existing performance analysis approaches used in process mining provide  solutions for the analysis of process completion times without resorting to simulation. The express technique relates mean times of the process activities to the mean execution time of the process. The full analysis technique constructs probability density functions (PDFs) of the overall process execution time based on PDFs of the process activities, thereby providing more detailed results.  The full analysis technique is based on a general model reduction approach that can be further extended and combined with different representations of PDFs. The time complexity of the reduction for discrete and continuous PDFs (in the form of Gaussian mixture models) is provided. The proposed approaches are implemented, \hl{compared to the simulation technique}, and tested on real-world event data demonstrating their applicability in practice. For the full analysis techniques, performance accuracy of the discovered models is additionally analyzed and discussed. \hl{The experimental results indicate that the discrete approach and simulation techniques offer better performance accuracy. However, the discrete approach is more time-efficient than the simulation for small support sizes. In contrast, the continuous approach enables the discovery of compact and interpretable process models.} As a future work, we plan to further enhance the implementation of the full analysis technique by combining continuous and discrete approaches. We also plan to extend the full analysis approach by applying queuing theory to estimate the overall execution time in the presence of queues.
\section{Appendix A. Formal proofs}
\label{sec:appendix}

\begin{customlemma}{1}[Properties of limiting probabilities of Markov process flows]
 \label{custom_lemma}
 Let $(\mathbf{P},s,e)$~be a Markov process flow and $\pi_r$ be a limiting probability for state $r$. Then $\pi_r=\pi_s\sum_{l=0}^\infty \widehat{p}_{s,r}^{\text{ }l}$, \ie the limiting probability of state $r$ can be defined as the limiting probability of the start state multiplied by the sum of all transition probabilities from $s$ to $r$ without transiting through $(e,s)$ edge.  
 \begin{proof}
 By \autoref{theor:lim}, $\pi_r=\sfrac{1}{d}\lim\limits_{n\to\infty}p_{s,r}^{dn}$, and this limit can be represented as: 
 $$ \pi_r=\frac{1}{d}\lim\limits_{{\substack{n_1\to\infty \\ n_2\to\infty}}}p_{s,s}^{dn_1}\sum\limits_{l=0}^{n_2} \text{  }\widehat{p}_{s,r}^{\text{ }l}\,, 
 $$ because if a path from $s$ to $r$ contains $e$, state $s$ will be visited once again, and, hence, we consider all the possibilities of reaching $r$ from $s$ in $dn_1+n_2$ steps in total. The limit as $n_1,n_2\to\infty$ equals $\pi_s\sum_{l=0}^\infty \widehat{p}_{s,r}^{\text{ }l}$, and, hence $\pi_r=\pi_s\sum_{l=0}^\infty \widehat{p}_{s,r}^{\text{ }l}$.
 \end{proof}
\end{customlemma}

\begin{customthm}{2}[Mean of process execution times]
\label{custom_theor}
Let $(\mathbf{P},\mathbf{F},s,e)$ be a semi-Markov process flow. Then, the mean time~$\mu$ of hitting time for state $e$ when the process has started in state $s$ (the mean of the overall execution time) can be calculated as follows: 
\begin{equation}
% \label{eqn:mean}
\mu=\frac{1}{\pi_s}\sum\limits_{{\substack{i=1, \\ i\neq e}}}^{m}\pi_i\mu_i \,,
\end{equation}
where $\pi_s$ is the limiting probability for the start state $s$, $\pi_i$ and $\mu_i$ are the limiting probability and the mean hitting time, respectively, for the state $i$, and $i$ iterates over all $m$ states except end state $e$.
\begin{proof}
Consider a matrix $\mathbf{Q}=(q_{i,j}(t))_{1\leq i,j\leq m}$, $t\in\mathbb{R}$, such that $q_{i,j}(t)=\widehat{p}_{i,j}f_{i,j}(t)$, where $\widehat{p}_{i,j}$ is the element of $\mathbf{\widehat{P}}$ (the~probability matrix corresponding to the original matrix $\mathbf{P}$, but with no edges from the state $e$). We define matrix $\mathbf{Q}^n$ as follows: $\mathbf{Q}^n=\mathbf{Q}^{n-1}*\mathbf{Q}$, where $n\geq 1$, $\mathbf{Q}^1=\mathbf{Q}$. Multiplication for  matrices of functions we define in such a way that $C=A*B$, when $c_{i,j}(t)=\sum_{r=1}^m a_{i,r}(t)*b_{r,j}(t)$, where $*$ is the convolution operation. By $q_{i,j}^n(t)$ we will denote the $i,j$-element of the matrix $\mathbf{Q}^n$.

As can be inferred from~\cite{10.1214/aoms/1177704864}, the overall process execution time (hitting state $e$ assuming that the process has started in state $s$) is $f(t)=\sum_{n=1}^\infty q_{s,e}^n(t)$, \ie the weighted time of hitting~$e$ in 1 step, 2 steps, etc.
Hence, the mean of process execution time can be calculated as:
\begin{align*}
    \mu[\sum\limits_{n=1}^\infty q_{s,e}^n(t)]=&\sum\limits_{n=1}^\infty \mu[q_{s,e}^n(t)]=\\&\mu[q_{s,e}^1(t)]+\mu[q_{s,e}^2(t)]+\mu[q_{s,e}^3(t)]+\dots
    =\\&\mu[p_{s,e}f_{s,e}(t)]+
    \mu[\sum\limits_{{\substack{i=1 \\ i\neq e}}}^m(p_{s,i}f_{s,i}(t)*p_{i,e}f_{i,e}(t))]+\\&\mu[\sum\limits_{{\substack{i,j=1 \\ i,j\neq e}}}^m(p_{s,i}f_{s,i}(t)*p_{i,j}f_{i,j}(t)*p_{j,e}f_{j,e})]+\dots \,.
\end{align*}
Since $\mu[f(t)*g(t)]=\mu[f(t)]+\mu[g(t)]$, this expression can be rewritten as: 
\begin{align*}
p_{s,e}(&\mu[f_{s,e}(t)])+\sum\limits_{{\substack{i=1 \\ i\neq e}}}^mp_{s,i}p_{i,e}(\mu[f_{s,i}(t)]+\mu[f_{i,e}(t)])\\&+\sum\limits_{{\substack{i,j=1 \\ i,j\neq e}}}^m p_{s,i}p_{i,j}p_{j,e}(\mu[f_{s,i}(t)]+\mu[f_{i,j}(t)]+\mu[f_{j,e}])+\dots \,.
\end{align*}
After rearrangement of terms we get:
$$\sum\limits_{{\substack{i=1 \\ i\neq e}}}^m\sum\limits_{l=0}^\infty\widehat{p}_{s,i}^{\text{ }l}\sum\limits_{l=0}^\infty\widehat{p}_{i,e}^{\text{ }l}\mu[\sum\limits_{j=1}^mp_{i,j}f_{i,j}(t)]\,.
$$
By~\autoref{custom_lemma} and because $\sum_{l=0}^\infty\widehat{p}_{i,e}^{\text{ }l}=1$ (the probability of eventually hitting state $e$), the expression can be represented as:
$$\sum\limits_{{\substack{i=1 \\ i\neq e}}}^m\frac{\pi_i}{\pi_s}\mu[\sum\limits_{j=1}^mp_{i,j}f_{i,j}(t)]=\frac{1}{\pi_s}\sum\limits_{{\substack{i=1 \\ i\neq e}}}^{m}\pi_i\mu_i\,,
$$ where $\mu_i$ is the mean time for state $i$. 

\end{proof}
\end{customthm}

\begin{customthm}{3}[Log and model mean times]
Let $L$ be an event log and $M=(\mathbf{P},\mathbf{F},s,e)$ be a semi-Markov process flow discovered from this log by \autoref{alg:stochastic:matrix:discovery}, with an arbitrary order $k$ set as a parameter. Then, the mean process execution times for the log and the model are the same, \ie $\mu_M=\mu_L$, where $\mu_M$ is the mean $e$ hitting time and $\mu_L$ is the mean duration of a trace for the event log $L$.  
\begin{proof}

The mean trace duration for event log $L$ can be represented as the sum of durations of all traces divided by the number of traces, \ie $\mu_L=\sum_{\langle e_1\dots e_n\rangle\in\overline{L}} (f_t(e_n)-f_t(e_1)) /|\overline{L}|$, where $\overline{L}$ is the trace representation of $L$. That equals to the sum of time differences between all neighboring events divided by the number of traces. Formally, the mean trace duration is defined as: $$\mu_L=\sum\limits_{\langle e_1\dots e_n\rangle\in\overline{L}} \left[\sum\limits_{j=1}^{n-1} (f_t(e_{j+1})-f_t(e_j))\right] /|\overline{L}|.$$
For each event $e_j$ from $L$, consider the activity name sequence\footnote{Without loss of generality, we assume that the length of the sequence is~$k$; however, according to \autoref{alg:stochastic:matrix:discovery}, there are also sequences with a length less than $k$.} $\langle f_a(e_{j-k+1}),\allowbreak f_a(e_{j-k+2}),\dots,f_a(e_{j})\rangle$ of length $k$ and rename it as $\langle a_1,a_2,\dots,a_k\rangle$. This sequence corresponds to a state in $M$ and can occur in $L$ multiple times. Let $\#\langle a_1,a_2,\dots,a_k\rangle$ be the number of occurrences of $\langle a_1,a_2,\dots,a_k\rangle$ in $L$ and let $\mu_j$ be the mean time spent by $M$ in the state $\langle a_1,a_2,\dots,a_k\rangle$. Because $\mu_j$ is the sum of transition times over all events $e_j$, such that $f_a(e_{j-k+1})=a_1, f_a(e_{j-k+2})=a_2,\dots, f_a(e_{j})=a_k$, divided by the number of occurrences of $\langle a_1,a_2,\dots,a_k\rangle$, the trace mean duration can be defined as the following sum $\sum_{\langle a_1,a_2,\dots,a_k\rangle \text{ in } L} \#\langle a_1,a_2,\dots,a_k\rangle\mu_j /|\overline{L}|$. The number of occurrences of the sequence $\langle a_1,a_2,\dots,a_k\rangle$ equals to the weighted sum of the number of occurrences of preceding  
sequences, \ie $\#\langle a_1,a_2,\dots,a_k\rangle=\sum_{j=1}^m p_{j,i}\#\langle a'_0,a_1,a_2,\dots,\allowbreak a_{k-1}\rangle$, proportionally to the frequencies of transitions  which connect the corresponding states (\autoref{alg:stochastic:matrix:discovery}). Hence, the occurrences of $k$-length sequences satisfy the system equations: $\pi_i=\sum_{j=1}^m p_{j,i}\pi_j$ (from \autoref{theor:lim}), and are proportional to the corresponding limiting probabilities of $M$, i.e. $\#\langle a_1,a_2,\dots,a_k\rangle= c\pi_i$, where $c$ is a coefficient. Similarly, the number of traces $|\overline{L}|$ can be calculated as the number of occurrences of the last events, and it is equal to $c\pi_e=c\pi_s$. Thus, $\mu_L=\sum_{j=1}^{m'} \sfrac{c\pi_j\mu_j} {c\pi_s}=\sfrac{1}{\pi_s}\sum_{j=1}^{m'} \pi_j\mu_j$, where $m'$ is the total number of distinct activity name sequences of the length $k$ in~$L$. Since $\mu_s=\mu_e=0$, $\mu_L=\sfrac{1}{\pi_s}\sum_{{\substack{j=1,  j\neq e}}}^{m}\pi_j\mu_j$, that is $\mu_M$ by \autoref{custom_theor}.
\end{proof}
\end{customthm}

\section{Acknowledgements}
\label{sec:acknowledgements}
The authors are grateful to Alex Hu for his contributions to the idea of state removal order.

\bibliographystyle{IEEEtran}
\bibliography{access}

\begin{thebibliography}{10}
\providecommand{\url}[1]{#1}
\csname url@samestyle\endcsname
\providecommand{\newblock}{\relax}
\providecommand{\bibinfo}[2]{#2}
\providecommand{\BIBentrySTDinterwordspacing}{\spaceskip=0pt\relax}
\providecommand{\BIBentryALTinterwordstretchfactor}{4}
\providecommand{\BIBentryALTinterwordspacing}{\spaceskip=\fontdimen2\font plus
\BIBentryALTinterwordstretchfactor\fontdimen3\font minus
  \fontdimen4\font\relax}
\providecommand{\BIBforeignlanguage}[2]{{%
\expandafter\ifx\csname l@#1\endcsname\relax
\typeout{** WARNING: IEEEtran.bst: No hyphenation pattern has been}%
\typeout{** loaded for the language `#1'. Using the pattern for}%
\typeout{** the default language instead.}%
\else
\language=\csname l@#1\endcsname
\fi
#2}}
\providecommand{\BIBdecl}{\relax}
\BIBdecl

\bibitem{Aalst16}
W.~M.~P. van~der Aalst, \emph{Process Mining: Data Science in Action}.\hskip
  1em plus 0.5em minus 0.4em\relax Springer, 2016.

\bibitem{solti-predict-simulation}
A.~Rogge-Solti and M.~Weske, ``Prediction of remaining service execution time
  using stochastic petri nets with arbitrary firing delays,'' in
  \emph{Service-Oriented Computing}, S.~Basu, C.~Pautasso, L.~Zhang, and X.~Fu,
  Eds.\hskip 1em plus 0.5em minus 0.4em\relax Berlin, Heidelberg: Springer
  Berlin Heidelberg, 2013, pp. 389--403.

\bibitem{ROGGESOLTI20151}
A.~{Rogge-Solti} and M.~Weske, ``Prediction of business process durations using
  non-{M}arkovian stochastic petri nets,'' \emph{Information Systems}, vol.~54,
  pp. 1--14, 2015.

\bibitem{vandenabeele2022enhancing}
J.~Vandenabeele, G.~Vermaut, J.~Peeperkorn, and J.~De~Weerdt, ``Enhancing
  stochastic petri net-based remaining time prediction using k-nearest
  neighbors,'' vol. 3167, 2022, Conference paper, p. 9 – 24.

\bibitem{ROZINAT2009305}
A.~Rozinat, R.~S. Mans, M.~Song, and W.~M.~P. {van der Aalst}, ``Discovering
  simulation models,'' \emph{Inf. Systems}, vol.~34, no.~3, pp. 305--327, 2009.

\bibitem{CAMARGO2020113284}
M.~Camargo, M.~Dumas, and O.~González-Rojas, ``Automated discovery of business
  process simulation models from event logs,'' \emph{Decision Support Systems},
  vol. 134, p. 113284, 2020.

\bibitem{VANDERAALST2011450}
W.~M.~P. {van der Aalst}, H.~Schonenberg, and M.~Song, ``Time prediction based
  on process mining,'' \emph{Information Systems}, vol.~36, no.~2, pp.
  450--475, 2011, special Issue: Semantic Integration of Data, Multimedia, and
  Services.

\bibitem{10.1007/978-3-642-13094-6_5}
W.~M.~P. van~der Aalst, M.~Pesic, and M.~Song, ``Beyond process mining: From
  the past to present and future,'' in \emph{Advanced Information Systems
  Engineering}, B.~Pernici, Ed.\hskip 1em plus 0.5em minus 0.4em\relax Berlin,
  Heidelberg: Springer Berlin Heidelberg, 2010, pp. 38--52.

\bibitem{context-aware}
F.~Folino, M.~Guarascio, and L.~Pontieri, ``Discovering context-aware models
  for predicting business process performances,'' in \emph{On the Move to
  Meaningful Internet Systems: OTM 2012}, R.~Meersman, H.~Panetto, T.~Dillon,
  S.~Rinderle-Ma, P.~Dadam, X.~Zhou, S.~Pearson, A.~Ferscha, S.~Bergamaschi,
  and I.~F. Cruz, Eds.\hskip 1em plus 0.5em minus 0.4em\relax Berlin,
  Heidelberg: Springer Berlin Heidelberg, 2012, pp. 287--304.

\bibitem{5f0e8dd04572478fb450eefe4211d69f}
A.~Rogge-Solti, W.~M.~P. {van der Aalst}, and M.~Weske,
  ``\BIBforeignlanguage{English}{Discovering stochastic {P}etri nets with
  arbitrary delay distributions from event logs},'' in
  \emph{\BIBforeignlanguage{English}{Business Process Management Workshops :
  BPM 2013 International Workshops}}.\hskip 1em plus 0.5em minus 0.4em\relax
  Springer, 2014, pp. 15--27.

\bibitem{10.4108/icst.valuetools.2011.245715}
N.~Anastasiou, T.-C. Horng, and W.~Knottenbelt, ``Deriving generalised
  stochastic {P}etri net performance models from high-precision location
  tracking data,'' in \emph{5th International ICST Conference on Performance
  Evaluation Methodologies and Tools}, 2012.

\bibitem{10.1007/978-3-319-15895-2_41}
B.~Carrera and J.-Y. Jung, ``Constructing probabilistic process models based on
  hidden {M}arkov models for resource allocation,'' in \emph{Business Process
  Management Workshops}.\hskip 1em plus 0.5em minus 0.4em\relax Springer, 2015,
  pp. 477--488.

\bibitem{7785764}
Y.~Kang and V.~Zadorozhny, ``Process discovery using classification tree hidden
  semi-{M}arkov model,'' in \emph{IRI}, 2016, pp. 361--368.

\bibitem{10.1007/978-3-319-98648-7_10}
A.~Augusto, A.~Armas-Cervantes, R.~Conforti, M.~Dumas, M.~La~Rosa, and
  D.~Reissner, ``Abstract-and-compare: A family of scalable precision measures
  for automated process discovery,'' in \emph{Business Process
  Management}.\hskip 1em plus 0.5em minus 0.4em\relax Cham: Springer
  International Publishing, 2018, pp. 158--175.

\bibitem{10.1007/978-3-030-72693-5_20}
A.~T. Burke, S.~J.~J. Leemans, and M.~T. Wynn, ``Stochastic process discovery
  by weight estimation,'' in \emph{Process Mining Workshops}, S.~Leemans and
  H.~Leopold, Eds.\hskip 1em plus 0.5em minus 0.4em\relax Cham: Springer
  International Publishing, 2021, pp. 260--272.

\bibitem{10.1007/978-3-030-76983-3_16}
A.~Burke, S.~J.~J. Leemans, and M.~T. Wynn, ``Discovering stochastic process
  models by reduction and abstraction,'' in \emph{Application and Theory of
  Petri Nets and Concurrency}, D.~Buchs and J.~Carmona, Eds.\hskip 1em plus
  0.5em minus 0.4em\relax Cham: Springer International Publishing, 2021, pp.
  312--336.

\bibitem{9230330}
A.~Polyvyanyy, A.~Moffat, and L.~García-Bañuelos, ``An entropic relevance
  measure for stochastic conformance checking in process mining,'' in
  \emph{2020 International Conference on Process Mining (ICPM)}, 2020, pp.
  97--104.

\bibitem{opt_disc}
S.~J.~J. Leemans, T.~Li, M.~Montali, and A.~Polyvyanyy, ``Stochastic process
  discovery: Can it be done optimally?'' in \emph{Advanced Information Systems
  Engineering}, G.~Guizzardi, F.~Santoro, H.~Mouratidis, and P.~Soffer,
  Eds.\hskip 1em plus 0.5em minus 0.4em\relax Cham: Springer Nature
  Switzerland, 2024, pp. 36--52.

\bibitem{gram_disc}
H.~Alkhammash, A.~Polyvyanyy, and A.~Moffat, ``Stochastic directly-follows
  process discovery using grammatical inference,'' in \emph{Advanced
  Information Systems Engineering}, G.~Guizzardi, F.~Santoro, H.~Mouratidis,
  and P.~Soffer, Eds.\hskip 1em plus 0.5em minus 0.4em\relax Cham: Springer
  Nature Switzerland, 2024, pp. 87--103.

\bibitem{Leemans2019}
S.~J.~J. Leemans, A.~F. Syring, and W.~M.~P. van~der Aalst, ``Earth movers'
  stochastic conformance checking,'' in \emph{Business Process Management
  Forum}, ser. Lecture Notes in Business Information Processing, vol.
  360.\hskip 1em plus 0.5em minus 0.4em\relax Springer, 2019, pp. 127--143.

\bibitem{10.1007/978-3-031-16171-1_7}
E.~Bogdanov, I.~Cohen, and A.~Gal, ``Conformance checking over stochastically
  known logs,'' in \emph{Business Process Management Forum}, C.~Di~Ciccio,
  R.~Dijkman, A.~del R{\'i}o~Ortega, and S.~Rinderle-Ma, Eds.\hskip 1em plus
  0.5em minus 0.4em\relax Cham: Springer International Publishing, 2022, pp.
  105--119.

\bibitem{10.1007/978-3-030-49435-3_14}
S.~J.~J. Leemans and A.~Polyvyanyy, ``Stochastic-aware conformance checking: An
  entropy-based approach,'' in \emph{Advanced Information Systems Engineering},
  S.~Dustdar, E.~Yu, C.~Salinesi, D.~Rieu, and V.~Pant, Eds.\hskip 1em plus
  0.5em minus 0.4em\relax Cham: Springer International Publishing, 2020, pp.
  217--233.

\bibitem{LEEMANS2023102197}
S.~J. Leemans and A.~Polyvyanyy, ``Stochastic-aware precision and recall
  measures for conformance checking in process mining,'' \emph{Information
  Systems}, vol. 115, p. 102197, 2023.

\bibitem{leemans2023enjoy}
S.~J. Leemans, F.~M. Maggi, and M.~Montali, ``Enjoy the silence: Analysis of
  stochastic petri nets with silent transitions,'' \emph{Information Systems},
  vol. 124, p. 102383, 2024.

\bibitem{9980707}
A.~T. Burke, S.~J. Leemans, M.~T. Wynn, W.~M. van~der Aalst, and A.~H. ter
  Hofstede, ``Stochastic process model-log quality dimensions: An experimental
  study,'' in \emph{2022 4th International Conference on Process Mining
  (ICPM)}, 2022, pp. 80--87.

\bibitem{wait_times}
M.~Ali, F.~Milani, and M.~Dumas, ``Data-driven identification and analysis of
  waiting times in business processes,'' \emph{Bus Inf Syst Eng}, 2024.

\bibitem{10.1007/978-3-031-34560-9_11}
K.~Lashkevich, F.~Milani, D.~Chapela-Campa, I.~Suvorau, and M.~Dumas, ``Why am
  {I} waiting? {D}ata-driven analysis of waiting times in business
  processes,'' in \emph{Advanced Information Systems Engineering}.\hskip 1em
  plus 0.5em minus 0.4em\relax Cham: Springer Nature Switzerland, 2023, pp.
  174--190.

\bibitem{DBLP:conf/bpm/DenisovFA18}
V.~Denisov, D.~Fahland, and W.~M.~P. van~der Aalst, ``Unbiased, fine-grained
  description of processes performance from event data,'' in \emph{Business
  Process Management - 16th International Conference, {BPM} 2018, Sydney, NSW,
  Australia, September 9-14, 2018, Proceedings}, ser. Lecture Notes in Computer
  Science, vol. 11080.\hskip 1em plus 0.5em minus 0.4em\relax Springer, 2018,
  pp. 139--157.

\bibitem{10.1007/978-3-642-36285-9_23}
A.~Adriansyah and J.~Buijs, ``Mining process performance from event logs,'' in
  \emph{Business Process Management Workshops}.\hskip 1em plus 0.5em minus
  0.4em\relax Springer Berlin Heidelberg, 2013, pp. 217--218.

\bibitem{10.1007/978-3-642-12186-9_15}
B.~van Dongen. and A.~Adriansyah, ``Process mining: Fuzzy clustering and
  performance visualization,'' in \emph{Business Process Management
  Workshops}.\hskip 1em plus 0.5em minus 0.4em\relax Springer Berlin
  Heidelberg, 2010, pp. 158--169.

\bibitem{10.1145/3202710.3203151}
M.~Leemans, W.~M.~P. van~der Aalst, and M.~G.~J. van~den Brand, ``Hierarchical
  performance analysis for process mining,'' in \emph{Proceedings of the 2018
  International Conference on Software and System Process}, ser. ICSSP
  '18.\hskip 1em plus 0.5em minus 0.4em\relax New York, NY, USA: Association
  for Computing Machinery, 2018, p. 96–105.

\bibitem{tree_performance}
S.~J. van Zelst, L.~F.~R. Santos, and W.~M.~P. van~der Aalst, ``Data-driven
  process performance measurement and prediction: A process-tree-based
  approach,'' in \emph{Intelligent Information Systems}, S.~Nurcan and
  A.~Korthaus, Eds.\hskip 1em plus 0.5em minus 0.4em\relax Cham: Springer
  International Publishing, 2021, pp. 73--81.

\bibitem{SENDEROVICH2019240}
A.~Senderovich, M.~Weidlich, and A.~Gal, ``Context-aware temporal network
  representation of event logs: Model and methods for process performance
  analysis,'' \emph{Information Systems}, vol.~84, pp. 240--254, 2019.

\bibitem{SENDEROVICH201896}
A.~Senderovich, A.~Shleyfman, M.~Weidlich, A.~Gal, and A.~Mandelbaum, ``To
  aggregate or to eliminate? optimal model simplification for improved process
  performance prediction,'' \emph{Information Systems}, vol.~78, pp. 96--111,
  2018.

\bibitem{math11245001}
S.~McClean and L.~Yang, ``Semi-markov models for process mining in smart
  homes,'' \emph{Mathematics}, vol.~11, no.~24, 2023.

\bibitem{piulachs2024semimarkovmultistatemodelingapproaches}
\BIBentryALTinterwordspacing
\hl{Xavier Piulachs and Klaus Langohr and Mireia Besalú and Natalia Pallarès
  and Jordi Carratalà and Cristian Tebé and Guadalupe Gómez Melis},
  ``\hl{Semi-Markov multistate modeling approaches for multicohort event
  history data},'' \hl{2024}. [Online]. Available:
  \url{https://arxiv.org/abs/2402.12027}
\BIBentrySTDinterwordspacing

\bibitem{doi:10.1191/0962280202SM276ra}
\hl{Per Kragh Andersen and Niels Keiding}, ``\hl{Multi-state models for event
  history analysis},'' \emph{\hl{Statistical Methods in Medical Research}},
  vol. \hl{11}, no. \hl{2}, pp. \hl{91--115}, \hl{2002}.

\bibitem{10.1093/biostatistics/kxaa030}
\hl{Titman, Andrew C and Putter, Hein}, ``\hl{General tests of the Markov
  property in multi-state models},'' \emph{\hl{Biostatistics}}, vol. \hl{23},
  no. \hl{2}, pp. \hl{380--396}, \hl{09} \hl{2020}.

\bibitem{Kim2003EfficientCF}
D.~S. Kim, H.-J. Moon, J.-H. Bahk, W.~H. Kwon, and Z.~J. Haas, ``Efficient
  computations for evaluating extended stochastic petri nets using algebraic
  operations,'' \emph{International Journal of Control Automation and Systems},
  vol.~1, pp. 431--443, 2003.

\bibitem{4505529}
A.~Csenki, ``Flowgraph models in reliability and finite automata,'' \emph{IEEE
  Transactions on Reliability}, vol.~57, no.~2, pp. 355--359, 2008.

\bibitem{10.1007/978-3-030-85172-9_5}
L.~Carnevali, M.~Paolieri, R.~Reali, and E.~Vicario, ``Compositional safe
  approximation of response time distribution of complex workflows,'' in
  \emph{Quantitative Evaluation of Systems}, A.~Abate and A.~Marin, Eds.\hskip
  1em plus 0.5em minus 0.4em\relax Cham: Springer International Publishing,
  2021, pp. 83--104.

\bibitem{Ross85}
S.~Ross, \emph{Introduction to Probability Models}, 3rd~ed.\hskip 1em plus
  0.5em minus 0.4em\relax Orlando, Florida, USA: Academic Press, 1985.

\bibitem{iso}
``Quality management systems — fundamentals and vocabulary,'' International
  Organization for Standardization, Standard, Sep. 2005.

\bibitem{10.1007/3-540-47961-9_37}
W.~M.~P. van~der Aalst, A.~Hirnschall, and E.~Verbeek, ``An alternative way to
  analyze workflow graphs,'' in \emph{Advanced Information Systems
  Engineering}.\hskip 1em plus 0.5em minus 0.4em\relax Springer Berlin
  Heidelberg, 2002, pp. 535--552.

\bibitem{DBLP:journals/corr/abs-1812-07334}
A.~Polyvyanyy, A.~Solti, M.~Weidlich, C.~{Di Ciccio}, and J.~Mendling,
  ``Monotone precision and recall measures for comparing executions and
  specifications of dynamic systems,'' \emph{ACM Trans. Softw. Eng. Methodol.},
  vol.~29, no.~3, 2020.

\bibitem{van2010process}
W.~M.~P. van~der Aalst, V.~Rubin, E.~Verbeek, B.~van Dongen, E.~Kindler, and
  C.~G{\"u}nther, ``Process mining: a two-step approach to balance between
  underfitting and overfitting,'' \emph{SoSyM}, vol.~9, no.~1, p.~87, 2010.

\bibitem{tonbeta166}
A.~Weijters, W.~Aalst, and A.~Medeiros, ``{Process Mining with the Heuristics
  Miner-algorithm},'' Eindhoven, The Netherlands, Tech. Rep., 2006.

\bibitem{10.2307/2241953}
A.~Barron and C.-H. Sheu, ``Approximation of density functions by sequences of
  exponential families,'' \emph{The Annals of Statistics}, vol.~19, no.~3, pp.
  1347--1369, 1991.

\bibitem{Wilson72numericalmethods}
G.~Wilson and A.~Wragg, ``Numerical methods for approximating continuous
  probability density functions, over $[0, \infty)$, using moments,'' 1972.

\bibitem{Ching2004HigherorderMC}
W.-K. Ching, E.~Fung, and M.~Ng, ``Higher‐order {M}arkov chain models for
  categorical data sequences,'' \emph{Naval Research Logistics}, vol.~51, 2004.

\bibitem{hopcroft2001introduction}
J.~Hopcroft, R.~Motwani, and J.~Ullman, \emph{Introduction to Automata Theory,
  Languages, and Computation}, ser. Addison-Wesley series in computer
  science.\hskip 1em plus 0.5em minus 0.4em\relax Addison-Wesley, 2001.

\bibitem{pm4py}
A.~Berti, S.~van Zelst, and W.~M.~P. van~der Aalst, ``Process mining for python
  (pm4py): Bridging the gap between process- and data science,'' in \emph{ICPM.
  Demo Track.}, vol. 2374, 2019, pp. 13--16.

\bibitem{WardSteeman2013}
\BIBentryALTinterwordspacing
W.~Steeman, ``Bpi challenge 2013, incidents,'' 2013. [Online]. Available:
  \url{https://data.4tu.nl/articles/dataset/BPI_Challenge_2013_incidents/12693914/1}
\BIBentrySTDinterwordspacing

\bibitem{https://doi.org/10.4121/uuid:3f422315-ed9d-4882-891f-e180b5b4feb5}
\BIBentryALTinterwordspacing
B.~van Dongen, ``Bpi challenge 2020: Domestic declarations,'' 2020. [Online].
  Available:
  \url{https://data.4tu.nl/articles/dataset/BPI_Challenge_2020_Domestic_Declarations/12692543/1}
\BIBentrySTDinterwordspacing

\bibitem{https://doi.org/10.4121/uuid:895b26fb-6f25-46eb-9e48-0dca26fcd030}
\BIBentryALTinterwordspacing
------, ``Bpi challenge 2020: Request for payment,'' 2020. [Online]. Available:
  \url{https://data.4tu.nl/articles/dataset/BPI_Challenge_2020_Request_For_Payment/12706886/1}
\BIBentrySTDinterwordspacing

\bibitem{pmlr-v35-daskalakis14}
C.~Daskalakis and G.~Kamath, ``Faster and sample near-optimal algorithms for
  proper learning mixtures of gaussians,'' in \emph{Proceedings of The 27th
  Conference on Learning Theory}, ser. Proceedings of Machine Learning
  Research, M.~F. Balcan, V.~Feldman, and C.~Szepesvári, Eds., vol.~35.\hskip
  1em plus 0.5em minus 0.4em\relax Barcelona, Spain: PMLR, 13--15 Jun 2014, pp.
  1183--1213.

\bibitem{DBLP:journals/corr/LiS15a}
J.~Z. Li and L.~Schmidt, ``A nearly optimal and agnostic algorithm for properly
  learning a mixture of k gaussians, for any constant k,'' \emph{CoRR}, vol.
  abs/1506.01367, 2015.

\bibitem{Kullback1951OnIA}
S.~Kullback and R.~A. Leibler, ``On information and sufficiency,'' \emph{Annals
  of Mathematical Statistics}, vol.~22, pp. 79--86, 1951.

\bibitem{10.1214/aoms/1177704864}
R.~Pyke, ``{M}arkov renewal processes with finitely many states,'' \emph{The
  Annals of Mathematical Statistics}, vol.~32, no.~4, pp. 1243--1259, 1961.

\end{thebibliography}

\vspace{-40pt}
\begin{IEEEbiography}[{\includegraphics[width=1in,height=1.25in,clip,keepaspectratio]{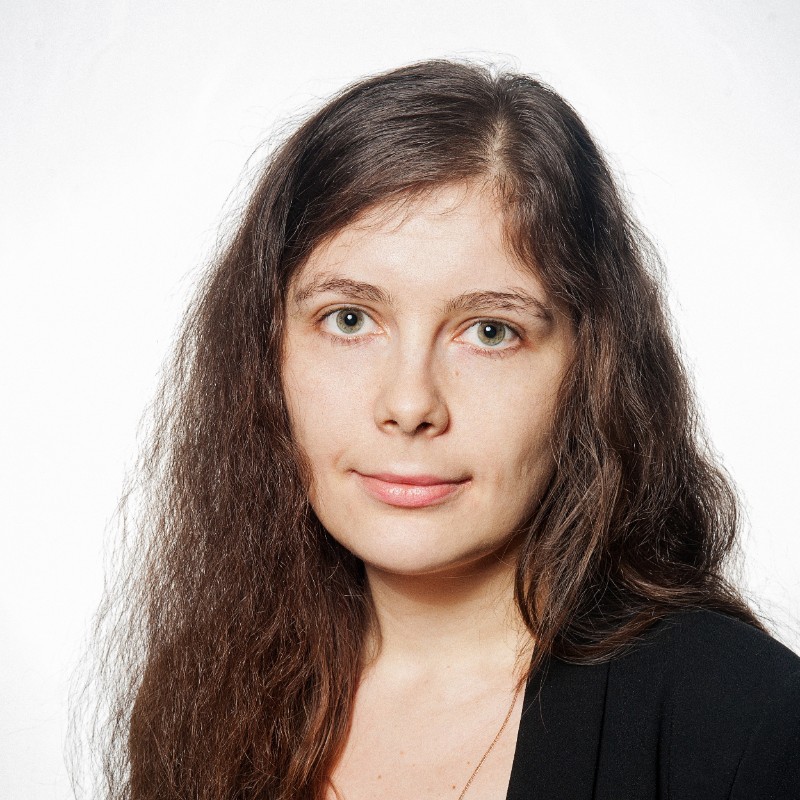}}]{Anna Kalenkova} received the Ph.D. degree in
process mining from the Eindhoven University of
Technology, Eindhoven, The Netherlands, in 2018,
under supervision of Prof. Wil van der Aalst and
Prof. Irina Lomazova. She is currently a Lecturer with the School of Computer and Mathematical Sciences, Faculty of Sciences, Engineering and Technology, The University of Adelaide,
Australia. 
Her interests include (process-oriented)
information systems, data analysis, business process modeling and process mining.
\end{IEEEbiography}

\vspace{-40pt}
\begin{IEEEbiography}[{\includegraphics[width=1in,height=1.25in,clip,keepaspectratio]{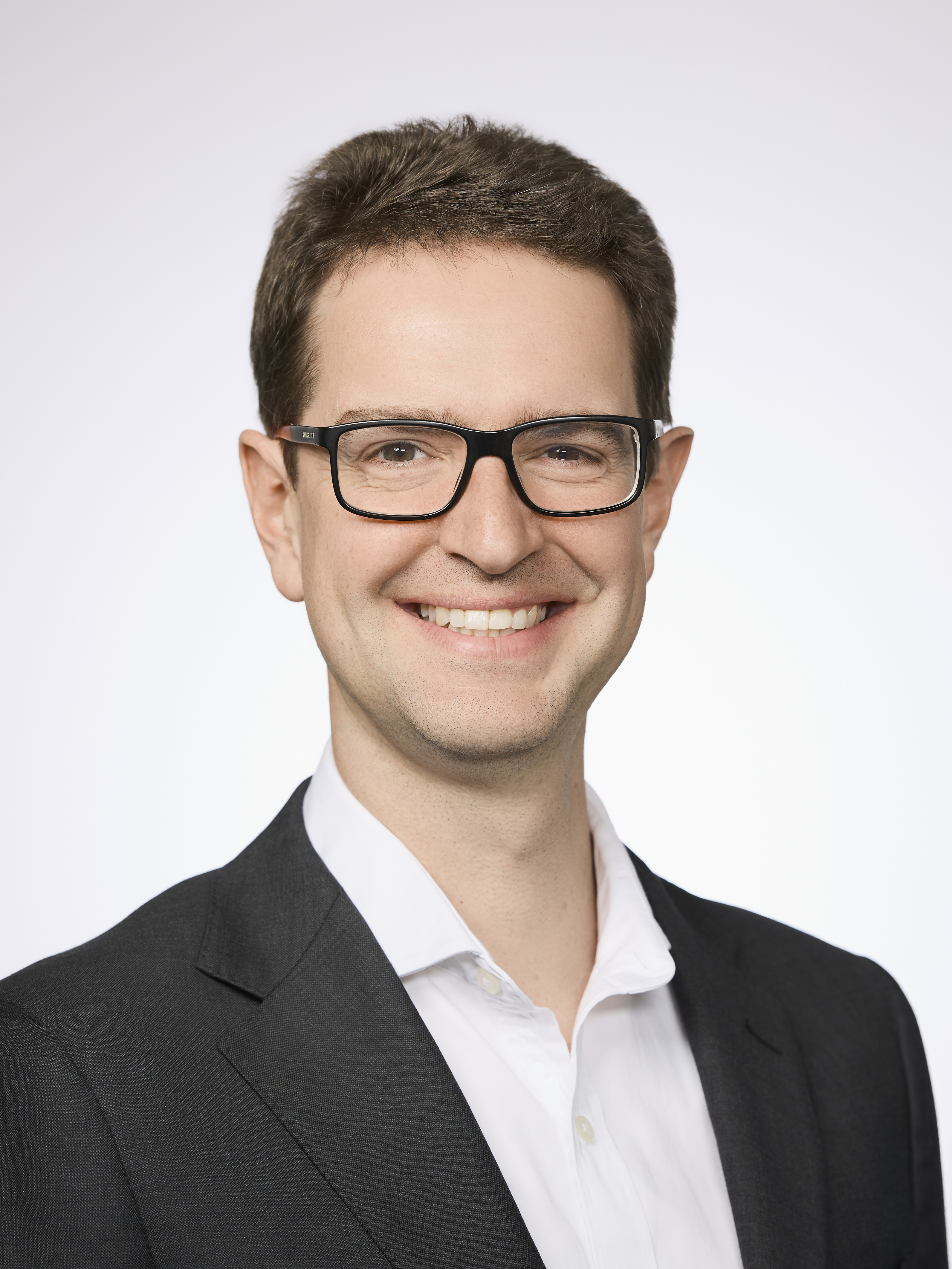}}]{Lewis Mitchell} received his Ph.D. degree in applied mathematics from University of Sydney in 2012. He is currently a Professor of Data Science at School of Computer and Mathematical Sciences, Faculty of Sciences, Engineering and Technology, the University of Adelaide, and a recipient of multiple awards, such as JH Michell Medal for outstanding new researchers, ANZIAM (2021), Australian Institute of Policy \& Science Young Tall Poppy Science Award (2018), ACEMS Outstanding Achievements Recognition Award (2018). His research interests are in mathematical modeling and data science, particularly in complex networks and   computational social science, human dynamics, and online social networks. 

\end{IEEEbiography}

\vspace{-40pt}
\begin{IEEEbiography}[{\includegraphics[width=1in,height=1.25in,clip,keepaspectratio,angle=270]{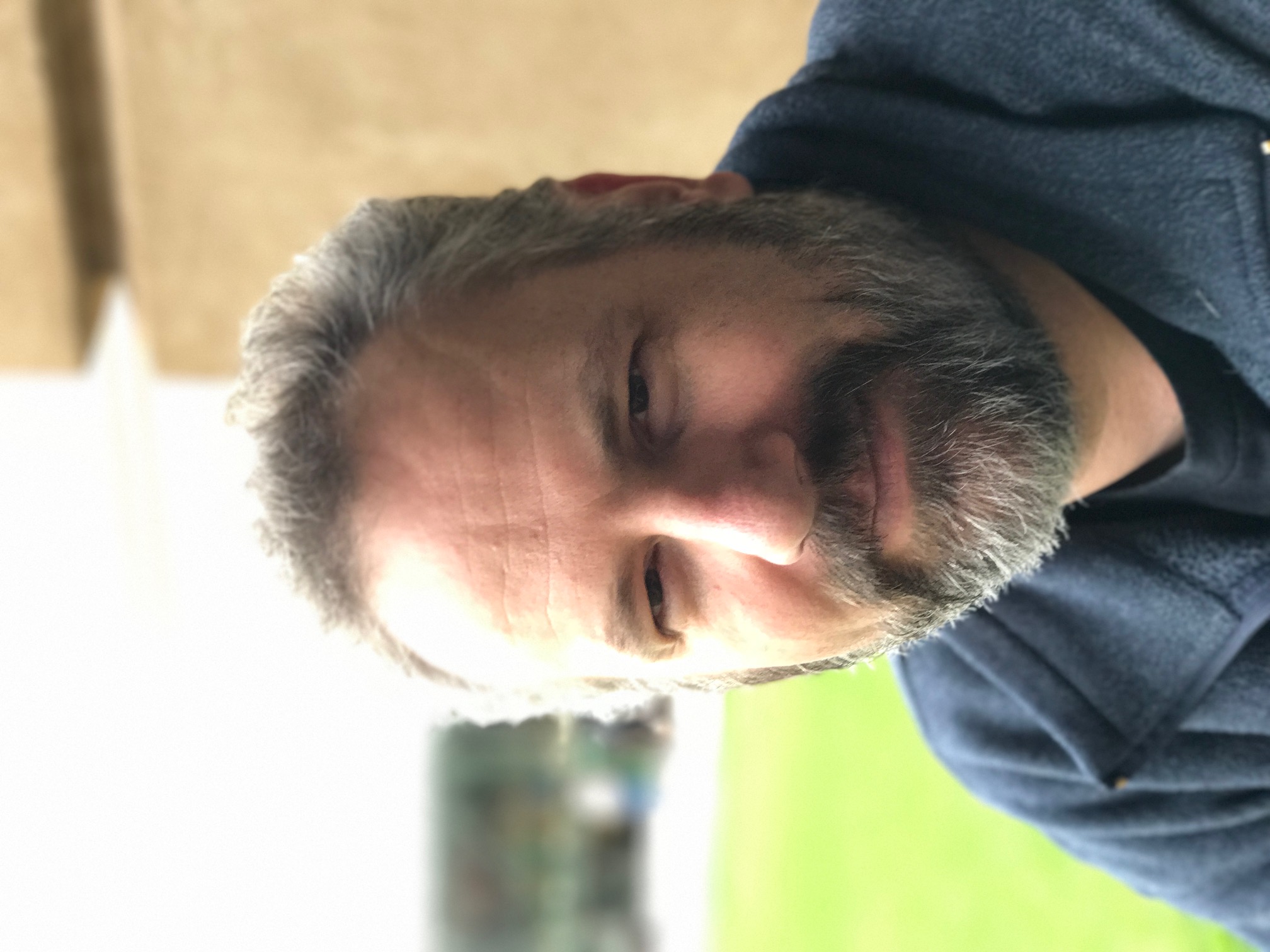}}]{Matthew Roughan} is a Professor at the University of Adelaide in the School of Computer and Mathematical Sciences, Australia. He was made a Fellow of the IEEE in 2019, and a Fellow of the ACM in 2018. He obtained his PhD in applied mathematics from the University of Adelaide in 1994. He has since worked for the Co-operative Research Centre for Sensor Signal and Information Processing (CSSIP), in conjunction with DSTO at the Software Engineering Research Centre at RMIT and the University of Melbourne, and in conjunction with Ericsson at the AT\&T Shannon Research Labs in the United States. His research interests are in traffic analysis, traffic anomaly detection, topology, performance analysis (including queueing theory),
routing (measurement and formal design), stochastic modelling, and signal processing.
\end{IEEEbiography}

\EOD

\end{document}